\documentclass[journal]{IEEEtran}
\pdfoutput=1
\newcommand{\SpaceBeforeSection}{\rule{1pt}{0pt}} %\vspace*{-\parskip} %Next subsection not properly separated with space by LaTeX

\usepackage{amsmath,amssymb}
\allowdisplaybreaks
\usepackage{mathrsfs} %Font used for sets
\usepackage{units}

\usepackage{color}
\definecolor{MyBlue}{rgb}{0,0.08,0.55}

\usepackage{graphicx} %Use option \usepackage[draft]{graphicx} to plot placeholders instead
\usepackage[tight,footnotesize]{subfigure} %For subfigures, use subfigure package, or newer version, subfig
\usepackage{array} %Extension of environments tabular and array
\usepackage[footnotesize]{caption} %Small captions [small], even smaller, [footnotesize], default is normalsize [normal]

\usepackage{url}

\def\hyph{-\penalty0\hskip0pt\relax}

\DeclareMathOperator*{\argmin}{arg\,min}
\DeclareMathOperator*{\argmax}{arg\,max}
\DeclareMathOperator{\oD}{D}                                            %Relative entropy
\DeclareMathOperator{\oE}{E}                                            %Expectation
\DeclareMathOperator{\oH}{H}                                            %Entropy
\DeclareMathOperator{\oI}{I}                                            %Mutual information and indicator function
\DeclareMathOperator{\oP}{P}                                            %Probability
\newcommand{\cD}{\mathcal{D}}                                           %Expected distortion
\newcommand{\cR}{\mathcal{R}}                                           %Expected risk or rate
\newcommand{\sC}{\mathscr{C}}                                           %Set C, quantization cell
\begin{document}

%SPACING FOR FORMULAS
%Use after \begin{document}, which calls the macro \normalsize and explicitly sets \abovedisplayskip
%IEEEtran settings
%\abovedisplayskip 1.5ex plus3pt minus1pt%
%\belowdisplayskip \abovedisplayskip%
%\abovedisplayshortskip 0pt plus3pt%
%\belowdisplayshortskip 1.5ex plus3pt minus1pt
%Overwrite settings
\setlength{\abovedisplayskip}{0.75ex plus2pt minus1pt}
\setlength{\belowdisplayskip}{0.75ex plus2pt minus1pt}
\setlength{\abovedisplayshortskip}{0pt plus 2pt}
\setlength{\belowdisplayshortskip}{0.75ex plus2pt minus1pt}

%%%%%%%%%%%%%%%%%%%%%%%%%%%%%%%%%%%%%%%%%%%%%%%%%%%%%%%%%%%%%%%%%%%%%%%%%%%%%%%%%%%%%%%%%%%%%%%%%%%%%%%%%%%%%%%%%%%%%%%%%%%%%%%%%%
%%%%%%%%%%%%%%%%%%%%%%%%%%%%%%%%%%%%%%%%%%%%%%%%%%%%%%%%%%%%%%%%%%%%%%%%%%%%%%%%%%%%%%%%%%%%%%%%%%%%%%%%%%%%%%%%%%%%%%%%%%%%%%%%%%
%TITLE AND AUTHORS
%%%%%%%%%%%%%%%%%%%%%%%%%%%%%%%%%%%%%%%%%%%%%%%%%%%%%%%%%%%%%%%%%%%%%%%%%%%%%%%%%%%%%%%%%%%%%%%%%%%%%%%%%%%%%%%%%%%%%%%%%%%%%%%%%%
%%%%%%%%%%%%%%%%%%%%%%%%%%%%%%%%%%%%%%%%%%%%%%%%%%%%%%%%%%%%%%%%%%%%%%%%%%%%%%%%%%%%%%%%%%%%%%%%%%%%%%%%%%%%%%%%%%%%%%%%%%%%%%%%%%

%111018: Claudia's affiliation
\title{On the Measurement of Privacy\\as an Attacker's Estimation Error}
\author{David~Rebollo-Monedero, Javier Parra-Arnau, Claudia Diaz and Jordi~Forn\'e% <-this % stops a space
\IEEEcompsocitemizethanks{\IEEEcompsocthanksitem D. Rebollo-Monedero, J. Parra-Arnau and J. Forn\'e are with
the Department of Telematics Engineering, Universitat Polit\`ecnica de Catalunya,
C.\ Jordi Girona 1-3, E-08034 Barcelona, Catalonia.\protect\\
E-mail: \{david.rebollo,javier.parra,jforne\}@entel.upc.edu.
\IEEEcompsocthanksitem C. Diaz is with the Katholieke Universiteit Leuven, ESAT/SCD/IBBT-COSIC,
Kasteelpark Arenberg 10, B-3001 Leuven-Heverlee, Belgium.\protect\\
E-mail: claudia.diaz@esat.kuleuven.be.}% <-this % stops a space
%\thanks{Manuscript prepared January, 2012.}} %prepared November, 2008; revised April, 2009; second revision July, 2009.}}
}
\maketitle

%%%%%%%%%%%%%%%%%%%%%%%%%%%%%%%%%%%%%%%%%%%%%%%%%%%%%%%%%%%%%%%%%%%%%%%%%%%%%%%%%%%%%%%%%%%%%%%%%%%%%%%%%%%%%%%%%%%%%%%%%%%%%%%%%%
%%%%%%%%%%%%%%%%%%%%%%%%%%%%%%%%%%%%%%%%%%%%%%%%%%%%%%%%%%%%%%%%%%%%%%%%%%%%%%%%%%%%%%%%%%%%%%%%%%%%%%%%%%%%%%%%%%%%%%%%%%%%%%%%%%
%ABSTRACT & KEYWORDS
%%%%%%%%%%%%%%%%%%%%%%%%%%%%%%%%%%%%%%%%%%%%%%%%%%%%%%%%%%%%%%%%%%%%%%%%%%%%%%%%%%%%%%%%%%%%%%%%%%%%%%%%%%%%%%%%%%%%%%%%%%%%%%%%%%
%%%%%%%%%%%%%%%%%%%%%%%%%%%%%%%%%%%%%%%%%%%%%%%%%%%%%%%%%%%%%%%%%%%%%%%%%%%%%%%%%%%%%%%%%%%%%%%%%%%%%%%%%%%%%%%%%%%%%%%%%%%%%%%%%%

\begin{abstract}

A wide variety of privacy metrics have been proposed in the literature
to evaluate the level of protection offered by privacy enhancing\hyph technologies.
Most of these metrics are specific to concrete systems and adversarial models, and are difficult to generalize or translate to other contexts.
Furthermore, a better understanding of the relationships between the different privacy metrics is needed to enable
more grounded and systematic approach to measuring privacy,
as well as to assist system designers in selecting the most appropriate metric for a given application.

% the paragraph below could still be improved (cdiaz)
%David, Javi 111018: Slightly revised
In this work we propose a theoretical framework for privacy\hyph preserving systems,
endowed with a general definition of privacy %that expresses
in terms of the estimation error incurred by an attacker who aims %at disclosing
to disclose the private information that the system is designed to conceal.
We show that our framework %allows the comparison of several well\hyph known metrics and their formulation
permits interpreting and comparing a number of well\hyph known metrics under a common perspective.
The arguments behind these interpretations are based on fundamental results related to the theories of information,
probability and Bayes decision.

\end{abstract}


%FOOTNOTES
\setcounter{footnote}{0} %Reset footnote counter unless same format is used
\renewcommand{\thefootnote}{(\alph{footnote})}  %If \alph,\arabic number style prererred to default

%%%%%%%%%%%%%%%%%%%%%%%%%%%%%%%%%%%%%%%%%%%%%%%%%%%%%%%%%%%%%%%%%%%%%%%%%%%%%%%%%%%%%%%%%%%%%%%%%%%%%%%%%%%%%%%%%%%%%%%%%%%%%%%%%%
%%%%%%%%%%%%%%%%%%%%%%%%%%%%%%%%%%%%%%%%%%%%%%%%%%%%%%%%%%%%%%%%%%%%%%%%%%%%%%%%%%%%%%%%%%%%%%%%%%%%%%%%%%%%%%%%%%%%%%%%%%%%%%%%%%
%1 INTRODUCTION
%%%%%%%%%%%%%%%%%%%%%%%%%%%%%%%%%%%%%%%%%%%%%%%%%%%%%%%%%%%%%%%%%%%%%%%%%%%%%%%%%%%%%%%%%%%%%%%%%%%%%%%%%%%%%%%%%%%%%%%%%%%%%%%%%%
%%%%%%%%%%%%%%%%%%%%%%%%%%%%%%%%%%%%%%%%%%%%%%%%%%%%%%%%%%%%%%%%%%%%%%%%%%%%%%%%%%%%%%%%%%%%%%%%%%%%%%%%%%%%%%%%%%%%%%%%%%%%%%%%%%

\section{Introduction}\label{sec:Introduction}
\noindent
The widespread use of information and communication technologies to conduct all kinds of activities has in recent years raised privacy concerns. There is a wide diversity of applications with a potential privacy impact, from social networking platforms to e-commerce or mobile phone applications.

% problem: ad hoc models and metrics
A variety of privacy\hyph enhancing technologies (PETs) have been proposed to enable the provision of new services and functionalities while mitigating potential privacy threats. The privacy concerns arising in different applications are diverse, and so are the corresponding privacy-enhanced solutions that address these concerns. Similarly, various ad hoc privacy metrics have been proposed in the literature to evaluate the effectiveness of PETs. The relationships between these different metrics have however not been investigated in depth, what leads to a fragmentation in the understanding of how privacy properties can be measured.

%David, Javi 111018: Minor rewording
% contributions
In this paper we consider a general, theoretical framework for privacy\hyph preserving systems and propose using the attacker's estimation error as privacy metric. We show that the most widely used privacy metrics, such as $k$\hyph anonymity, $l$\hyph diversity, $t$\hyph closeness, $\epsilon$\hyph differential privacy, as well as information\hyph theoretic metrics such as Shannon's entropy, min\hyph entropy, or mutual information, may be construed as %are in fact
particular cases of the estimation error.

%[Say something about optimization, how the metric/framework is useful to optimize the tradeoff cost-privacy.]
Privacy metrics, accompanied with utility metrics, provide a quantitative means of comparing the suitability of two or more privacy\hyph enhancing mechanisms, in terms of the privacy\hyph utility trade\hyph off posed.
Ultimately, such metrics will enable us to systematically build privacy\hyph aware information systems
by formulating design decisions as optimization problems, solvable theoretically or numerically,
capitalizing on a rich variety of mature ideas and powerful techniques from the wide field of optimization engineering.
%Having stressed the crucial importance of the measurement of privacy,
%we believe that our study into this matter may prove useful in two ways.

We illustrate how the general model can be instantiated in three very different areas of application, namely statistical disclosure control, anonymous communications and location\hyph based services.
% -------------------------------------------------------------------------------------- SDC --------------------------------------------------------------------------------------------
Statistical disclosure control (SDC)~\cite{Willenborg01B} is the research area that deals with the inherent compromise between protecting the privacy of the individuals in a microdata set and ensuring that those data are still useful for researchers. Traditionally, institutes and governmental statistical agencies have systematically gathered information about individuals with the aim of distributing those data to the research community~\cite{Jabine93OS}. However, the distribution of this information should not compromise respondents' privacy in the sense of revealing information about specific individuals.
% 121002 Claudia: my impression is that [4], [5], [6], [7] have been developed by researchers, rather than statistical agencies. rephrase?
% For this purpose, statistical agencies have studied and developed numerous mechanisms~\cite{Citteur93OS} and algorithms~\cite{Domingo02KDE,Domingo05DMKD,Solanas06COMPSTAT,Rebollo09IWDC} to apply to the microdata sets before releasing them.
Motivated by this, considerable research effort has been devoted to the development of privacy\hyph protecting mechanisms~\cite{Citteur93OS,Domingo02KDE,Domingo05DMKD,Solanas06COMPSTAT,Rebollo09IWDC}
to be applied to the microdata sets before their release.
In essence, these mechanisms rely upon some form of perturbation that permits enhancing privacy to a certain extent,
at the cost of losing some of the data utility with respect to the unperturbed version.

%A great effort has been devoted to the investigation of privacy metrics in the scenario of SDC.
With the aim of assessing the effectiveness of such mechanisms, numerous privacy metrics have been investigated.
Probably, the best\hyph known privacy metric is $k$\hyph \emph{anonymity}, which was first proposed in~\cite{Sweeney02UFKBS,Samarati01KDE}. In an attempt to address the limitations of this proposal, various extensions and enhancements were introduced later in~\cite{Truta06PDM,Machanavajjhala06ICDE,Li07ICDE,Brickell08KDD,Dwork06A,Rebollo10KDE}.
% 121002 Claudia: I would rephrase the phrase below, and remove the ref to Solove [16]". We are _not_ exploring the connection between "social, legal, and policy" conceptions of privacy and "technological" conceptions. Rather, we are generalizing privacy metrics that are very much technology-oriented.
%While all these proposals have contributed to some extent to a better comprehension of a concept that, for some, is virtually impossible to define~\cite{Solove09B},
%the SDC research community would undoubtedly benefit from the existence of a rule that could help them decide which privacy metric is the most appropriate for a particular application.
While all these proposals have contributed to some extent to the understanding of the privacy requirements of this field,
the SDC research community would undoubtedly benefit from the existence of a rule that could help them decide which privacy metric is the most appropriate for a particular application.
In other words, there is a need for the establishment of a framework that enables us to compare those metrics and to formulate them by using a common, general definition of privacy.

% ---------------------------------------------------------------------------- Anonymous communications-----------------------------------------------------------------------------
%In anonymous communications, the goal is to conceal who talks to whom against an adversary who observes the inputs and outputs of the anonymous communication channel.
% 121001 Claudia: Not strictly true. Often the goal is to conceal the sender from the recipient. Furthermore, low-latency anon comm (crowds, tor) do _not_ provide any privacy towards this adversary. Perhaps rephrase to "one of the goals of anonymous communications systems" ?
In anonymous communications, one of the goals is to conceal who talks to whom against an adversary who observes the inputs and outputs of the anonymous communication channel.
Mixes~\cite{Chaum81CACM,Cottrell94,Danezis03SP} are a basic building block for implementing anonymous communication channels. Mixes perform cryptographic operations on messages such that it is not possible to correlate their inputs and outputs based on their bit patterns. In addition, mixes delay and reorder messages to hinder the linking of inputs and outputs based on timing information. Delaying messages has an impact on the usability of the system, and therefore imposes a cost on the system. On the other hand, higher delays allow for stronger levels of privacy. There is thus a trade\hyph off between delay (cost) and anonymity (privacy), and optimizing the level of anonymity for a given expected delay is interesting to extract as much protection as possible from the anonymous channel at the lower possible cost.

% ------------------------------------------------------------------------------ LBS ------------------------------------------------------------------------------
In the end, we approach the particularly rich, important example of location\hyph based services (LBSs),
where users submit queries along with the location to which those queries refer.
An example would be the query ``Where is the nearest Italian restaurant?'', accompanied by the geographic coordinates of the user's current location.
% 121002 Claudia: not sure the distinction between TTP-based and collaborative architectures is relevant, and I think it introduces confusion. I would compress this paragraph.
%In this scenario, one of the conceptually simplest approaches to preserve user privacy consists in including a trusted third party (TTP) acting as an intermediary between the user and the information service provider, what effectively hides the identity of the user.
%Additional solutions have been proposed, many of them based on an intelligent perturbation of the user coordinates submitted to the provider~\cite{Duckham01CEUS}.
In this scenario, a wide range of approaches have been proposed,
many of them based on an intelligent perturbation of the user coordinates submitted to the provider~\cite{Duckham01CEUS}.
Essentially, users may contact an \emph{untrusted} LBS provider directly, perturbing their location information so as to
hinder providers in their efforts to compromise user privacy in terms of location,
although clearly not in terms of query contents and activity,
and at the cost of an inaccurate answer.
In a nutshell, this approach presents again the inherent trade\hyph off between data utility and privacy common to any perturbative privacy method.

% ------------------------------------------------------------------------------ Summary ------------------------------------------------------------------------------
The survey of privacy metrics, the detailed analysis of their connection with information theory,
and the mathematical unification as an attacker's estimation error presented in this paper
shed new light on the understanding of those metrics and their suitability when it comes to applying them to specific scenarios.
In regard to this aspect, two sections are devoted to the classification of several privacy metrics,
showing the relationships with our proposal and the correspondence with assumptions on the attacker's strategy.
While the former section approaches this from a theoretical perspective, %the latter is written as a guide to help system designers choose the appropriate metrics,
the latter illustrates the applicability of our framework to help system designers choose the appropriate metrics,
without having to delve into the mathematical details.
%In regard to this aspect, an entire section is specially devoted to a classification of several privacy metrics,
%written as a guide to help system designers choose the appropriate ones,
%showing the relationships with others, and the correspondence with assumptions on the attacker's strategy.
We also hope to illustrate the riveting intersection between the fields of information privacy and information theory,
in an attempt towards bridging the gap between the respective communities.
Moreover, the fact that our metric boils down to an estimation error opens the possibility of applying notions and results from the mature, vast field of estimation theory~\cite{Lehmann83B}.

%and the potential for each community to benefit from the expertise of the other.

%%%%%%%%%%%%%%%%%%%%%%%%%%%%%%%%%%%%%%%%%%%%%%%%%%%%%%%%%%%%%%%%%%%%%%%%%%%%%%%%%%%%%%%%%%%%%%%%%%%%%%%%%%%%%%%%%%%%%%%%%%%%%%%%%%
%%%%%%%%%%%%%%%%%%%%%%%%%%%%%%%%%%%%%%%%%%%%%%%%%%%%%%%%%%%%%%%%%%%%%%%%%%%%%%%%%%%%%%%%%%%%%%%%%%%%%%%%%%%%%%%%%%%%%%%%%%%%%%%%%%
%2 BACKGROUND
%%%%%%%%%%%%%%%%%%%%%%%%%%%%%%%%%%%%%%%%%%%%%%%%%%%%%%%%%%%%%%%%%%%%%%%%%%%%%%%%%%%%%%%%%%%%%%%%%%%%%%%%%%%%%%%%%%%%%%%%%%%%%%%%%%
%%%%%%%%%%%%%%%%%%%%%%%%%%%%%%%%%%%%%%%%%%%%%%%%%%%%%%%%%%%%%%%%%%%%%%%%%%%%%%%%%%%%%%%%%%%%%%%%%%%%%%%%%%%%%%%%%%%%%%%%%%%%%%%%%%
\section{Related work}\label{sec:Background}
\noindent
In this section we provide an overview of privacy metrics with an emphasis on those used in the three applications under study: anonymous communications, location\hyph based services and statistical disclosure control.

%%%%%%%%%%%%%%%%%%%%%%%%%%%%%%%%%%%%%%%%%%%%%%%%%%%%%%%%%%%%%%%%%%%%%%%%%%%%%%%%%%%%%%%%%%%%%%%%%%%%%%%%%%%%%%%%%%%%%%%%%%%%%%%%%%
%2.1 ANONYMOUS COMMUNICATION SYSTEMS
%%%%%%%%%%%%%%%%%%%%%%%%%%%%%%%%%%%%%%%%%%%%%%%%%%%%%%%%%%%%%%%%%%%%%%%%%%%%%%%%%%%%%%%%%%%%%%%%%%%%%%%%%%%%%%%%%%%%%%%%%%%%%%%%%%
\subsection{Anonymous\hyph Communication Systems and Location\hyph Based Services }\label{sec:Background:ACS_LBS}
\noindent
%\textbf{CLAUDIA'S PART GOES HERE}.
%[1 paragraph ]
% mixes: chaum, mixmaster, mixminion
Mixes were proposed by Chaum~\cite{Chaum81CACM} in 1981, and are a basic building block for implementing high-latency anonymous communications. A mix takes a number of input messages, and outputs them in such a way that it is infeasible to link an output to its corresponding input. In order to achieve this goal, the mix changes the appearance (by encrypting and padding messages) and the flow of messages (by delaying and reordering them). Mixmaster~\cite{Cottrell94} and Mixminion~\cite{Danezis03SP} are more advanced versions of the Chaumian mix~\cite{Chaum81CACM}, and they haven been deployed to provide anonymous email services.

% low-latency: OR, Tor, Crowds ??

% pre-info-theory metrics
Several metrics have been proposed in the literature to assess the level of anonymity provided by anonymous\hyph communication systems (ACSs). Reiter and Rubin~\cite{Reiter98ISS} define the degree of anonymity as a probability $1-p$, where $p$ is the probability assigned by an attacker to the potential initiators of a communication. In this model, users are more anonymous as they appear (towards a certain adversary) to be less likely of having sent a message, and the metric is thus computed individually for each user and for each communication. Berthold et al.~\cite{Berthold00WDIAU} on the other hand define the degree of anonymity as
%$A = \log_2 n$, where $n$ is
%David 111018: unnecessary math removed, explicit mention of Hartley entropy
the binary logarithm of the number of users of the system, which may be regarded as a Hartley entropy.
This metric only depends on the number of users of the system, and does not take into account that some users might appear as more likely senders of a message than others.

% shannon entropy
Information theoretic anonymity metrics were independently proposed in two papers. The metric proposed by Serjantov and Danezis~\cite{Serjantov02PET} uses Shannon's entropy as measure of the effective anonymity set size. The metric proposed by Diaz et
%111018 Javi: English correction
al.~\cite{Diaz02PET} normalizes Shannon's entropy to obtain a degree of anonymity on a scale from 0 to~1.

% other entropies
Toth et al.~\cite{Toth2004NWSITS} argue that Shannon entropy may not provide relevant information to some users, as it considers the average instead of the worst-case scenario for a particular user. They suggest using instead a local anonymity measure computed from min-entropy and max-entropy. Clauss and Schiffner~\cite{Clauss06WDIM} proposed R\'enyi entropy as a generalization of Shannon, min- and max-entropy-based anonymity metrics.

%The intuition of measuring privacy as an attacker's estimation error, which we attempt to unravel in this paper,
%is touched upon in~\cite{Shokri09WPES,Shokri11SP}. In the context of location privacy for mobile wireless networks, the cited papers formalize a privacy metric specially tailored to this location application, essentially an expected distortion in the estimation of a user's trajectory by an adversary. The proposed metric is then compared to existing, related metrics, also specific to location privacy.

% non info-theory
Other anonymity metrics in the literature include possibilistic (instead of probabilistic) approaches, such as those proposed by Syverson and Stubblebine~\cite{Syverson99FM}, Mauw et al.~\cite{Mauw04ESORICS}, or Feigenbaum et al.~\cite{Feigenbaum07FC}. According to these metrics, subjects are considered anonymous if the adversary cannot determine their actions with absolute certainty. Finally, Edman et al.~\cite{Edman07ISI} propose a combinatorial anonymity metric that measures the amount of information needed to reveal the full set of relationships between the inputs and the outputs of a mix. Some extensions of this model were proposed by Gierlichs et al.~\cite{Gierlichs08WPES} and by Bagai et al.~\cite{Bagai11PET}.

Having examined the most relevant metrics in the field of anonymous communications, now we briefly touch upon some of the proposals intended for the scenario of LBS.
Particularly, the issue of quantifying privacy in this scenario has been explored in~\cite{Shokri09WPES} and revisited shortly afterwards in~\cite{Shokri11SP}. At a conceptual level, we encounter the same underlying principle proposed here, in the sense that the authors propose to measure privacy as the adversary's expected estimation error for that particular context. We shall discuss later in Sec.~\ref{sec:Numerical:LBS} that their specific metric for LBS may be construed as an illustrative special case of our own work, and describe notable differences with respect to our generic framework.

%%%%%%%%%%%%%%%%%%%%%%%%%%%%%%%%%%%%%%%%%%%%%%%%%%%%%%%%%%%%%%%%%%%%%%%%%%%%%%%%%%%%%%%%%%%%%%%%%%%%%%%%%%%%%%%%%%%%%%%%%%%%%%%%%%
%2.2 PRIVACY CRITERIA IN STATISTICAL DISCLOSURE CONTROL
%%%%%%%%%%%%%%%%%%%%%%%%%%%%%%%%%%%%%%%%%%%%%%%%%%%%%%%%%%%%%%%%%%%%%%%%%%%%%%%%%%%%%%%%%%%%%%%%%%%%%%%%%%%%%%%%%%%%%%%%%%%%%%%%%%
\subsection{Privacy Criteria in Statistical Disclosure Control}\label{sec:Background:SDC}
\noindent
%In Sec.~\ref{sec:Introduction}, we shortly commented on the privacy risks derived from the dissemination of microdata sets.
In statistical disclosure control terminology, a microdata set is a database whose records contain information at the level of individual respondents.
In those databases, each row corresponds to an individual and each column, to an attribute. According to the nature of attributes, we may classify them into
\emph{identifiers}, \emph{key attributes} or \emph{quasi}\hyph\emph{identifiers}, or \emph{confidential attributes}.
On the one hand, identifiers allow to unequivocally identify individuals. For example, it would be the case of social security numbers or full names, which would be removed before the publication of the microdata set.
On the other hand, key attributes are those attributes that, in combination, may be linked with external information to reidentify the respondent to whom the records in the
microdata set refer. Last but not least, confidential attributes contain sensitive information on the respondents, such as health condition, political affiliation, religion or salary.

\begin{figure}
\centering\hspace*{\fill}
\subfigure[Original data]%
{\includegraphics[scale=0.53]{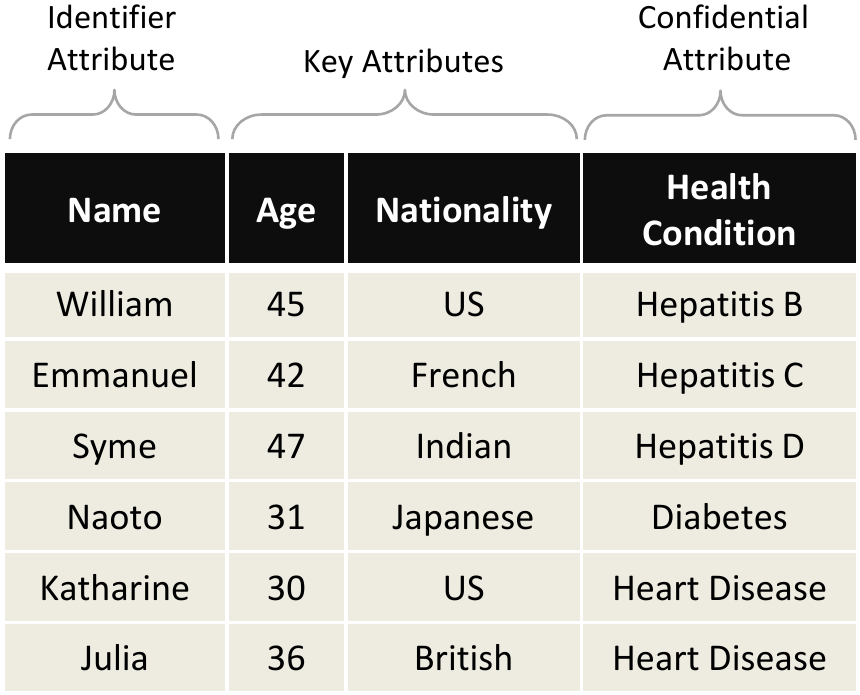}%
\label{fig:original_data}}\hfill
\subfigure[Perturbed data]%
{\includegraphics[scale=0.53]{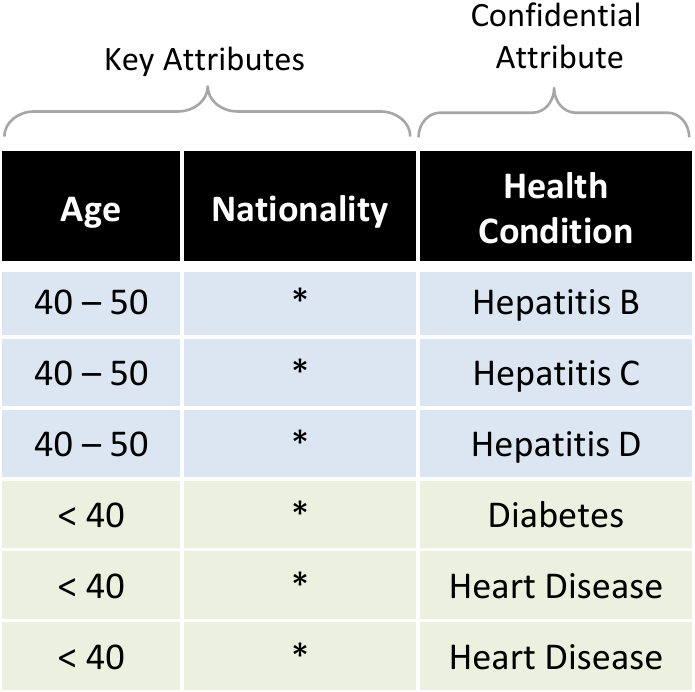}%
\label{fig:released_data}}\hspace*{\fill}
\caption{
%We apply the mechanisms of generalization and suppression to the key attributes ``age'' and ``nationality'', respectively.
%As a result, the released table~(b) satisfies the requirement of 3\hyph anonymity,
%which means that each tuple of key attributes in this table is shared by at least 3 records.}
We apply the mechanisms of generalization and suppression to the key attributes ``age" and ``nationality" respectively,
in such a manner that the requirement of 3\hyph anonymity is satisfied.
The upshot of such perturbation is that each tuple of key attributes in the released table (b) is shared by at least 3 records.
This means that an attacker who knows the key attribute values of a particular respondent cannot ascertain the record of this respondent beyond a subgroup of 3 records in any public database with identifier attributes.}
\label{fig:generalization_and_suppression}
\end{figure}

$k$\hyph\emph{Anonymity}~\cite{Samarati01KDE,Sweeney02UFKBS} is the requirement that each tuple of key attribute values be shared by at least $k$~records in the database. This condition is illustrated in Fig.~\ref{fig:generalization_and_suppression}, where a microdata set is $k$\hyph anonymized before publishing it.
Particularly, this privacy criterion is enforced by using generalization and suppression, two mechanisms by which key attribute values are respectively coarsened and eliminated. As a result, all key attribute values within each group are replaced by a common tuple, and thus a record cannot be unambiguously linked to any public database containing identifiers. Consequently, $k$\hyph anonymity is said to protect microdata against \emph{linking attacks}.

Unfortunately, while this criterion prevents identity disclosure, it may fail against the disclosure of the confidential attribute.
Concretely, suppose that a privacy attacker knows Emmanuel's key attribute values.
If the attacker learns that he is included in the released table depicted in Fig.~\ref{fig:released_data},
then the attacker may conclude that this patient suffers from hepatitis even though the attacker is unable to ascertain which record belongs to this individual.
This is known as~\emph{similarity attack},
meaning that values of confidential attributes may still be semantically similar.
More generally, the \emph{skewness attack} exploits the difference between
the prior distribution of confidential attributes in the whole data set
and the posterior distribution of those attributes within a specific group.

All these vulnerabilities motivated the appearance of a number of proposals, some of which we now overview.
An enhancement of $k$\hyph anonymity called $p$\hyph sensitive $k$\hyph anonymity~\cite{Truta06PDM}
incorporates the additional restriction that there be at least $p$ distinct values
for each confidential attribute within each $k$\hyph anonymous group.
With the aim of addressing the data utility loss incurred by large values of~$p$,
$l$\hyph diversity~\cite{Machanavajjhala06ICDE} proposes instead that there be at least $l$ ``well\hyph represented'' values for each confidential attribute.
Unfortunately, both proposals are still vulnerable to similarity attacks and skewness attacks.
% JAVI - ÍDEM
%These vulnerabilities motivated the appearance of a number of proposals, some of which we now overview. An enhancement of \emph{k}\hyph anonymity called
%\emph{p}\hyph sensitive \emph{k}\hyph anonymity was presented in~\cite{Truta06PDM}. This criterion incorporates the additional restriction that there be at least \emph{p} distinct values for each confidential attribute within the group of records sharing the same tuple of perturbed key attribute values. With the aim of addressing the data utility loss incurred by large values of~\emph{p}, \emph{l}\hyph diversity~\cite{Machanavajjhala06ICDE} proposes instead that there be at least $l$~``well\hyph represented'' values for each confidential attribute.
%Lamentably, both of them
%are vulnerable to \emph{similarity attacks} in the sense that values of confidential attributes may still be semantically similar within a group of records with the same tuple of perturbed key attribute values. Furthermore, \emph{l}\hyph diversity is susceptible against the exploitation of the difference between the prior distribution of confidential attributes in the whole data set and the posterior
%distribution of those attributes within a specific group. This is normally referred to as \emph{skewness attack}.

In an attempt to overcome all these deficiencies, \emph{t}\hyph \emph{closeness}~\cite{Li07ICDE} was proposed. A microdata set satisfies $t$\hyph closeness if, for each group of records with the same tuple of perturbed key attribute values,
a measure of discrepancy between the posterior and prior distributions does not exceed a threshold~$t$.
Inspired by this measure, \cite{Rebollo10KDE}~defines an \emph{(average) privacy risk} as the conditional
\emph{Kullback\hyph Leibler (KL) divergence} between the posterior and the prior distributions,
a measure that may be regarded as an averaged version of $t$\hyph closeness.
Further, this average privacy risk is shown to be equal to the mutual information between the confidential attributes
and the observed, perturbed key attributes, and, finally, a connection is established with Shannon's rate\hyph distortion theory.
A related criterion named $\delta$\hyph \emph{disclosure} is proposed in~\cite{Brickell08KDD},
a yet stricter version that measures the maximum absolute log ratio between the prior and the posterior distributions.
Lastly, \cite{Dwork06A}~analyzes privacy for interactive databases,
where a randomized perturbation rule is applied to a true answer to a query, before returning it to the user.
Consider two databases that differ only by one record, but are subject to a common perturbation rule.
Conceptually, the randomized perturbation rule is said to satisfy the
$\epsilon$\hyph \emph{differential privacy} criterion if the two corresponding probability distributions of the perturbed answers
are similar, according to a certain inequality.
% 121001 Claudia: ambiguous: which "certain inequality"?
% 121001 Javi to Claudia: I agree with you. But since notation has not been introduced yet, I think it would be better to leave it as it is. I changed next line to say we will provide more details on about this.
Later in Sec.~\ref{sec:Theory:noHamming} we provide further details about these privacy criteria and relate them in terms of our formulation.

%%%%%%%%%%%%%%%%%%%%%%%%%%%%%%%%%%%%%%%%%%%%%%%%%%%%%%%%%%%%%%%%%%%%%%%%%%%%%%%%%%%%%%%%%%%%%%%%%%%%%%%%%%%%%%%%%%%%%%%%%%%%%%%%%%
%3 Preliminaries
%%%%%%%%%%%%%%%%%%%%%%%%%%%%%%%%%%%%%%%%%%%%%%%%%%%%%%%%%%%%%%%%%%%%%%%%%%%%%%%%%%%%%%%%%%%%%%%%%%%%%%%%%%%%%%%%%%%%%%%%%%%%%%%%%%
\section{Preliminaries}\label{sec:Background:BDT}
\noindent
%Overview
In this section, we shall present our convention regarding random variables (r.v.'s) and probability distributions.
Next, we shall introduce some elementary concepts for those readers who are not familiar with Bayes decision theory (BDT).

% Convention with regard to random variables and probability distributions.
Throughout this paper, we shall follow the convention of using uppercase letters to denote r.v.'s, and lowercase letters to the
particular values they take on. We shall call~\emph{alphabet} the values an r.v.\ takes on.
Probability mass functions (PMFs) are denoted by~$p$, subindexed by the corresponding r.v.
%\ in case of ambiguity risk. For example, both $p_X(x)$ and $p(x)$ denote the value of the function $p_X$ at~$x$, which aids in writing more concise equations.
Accordingly, $p_X(x)$ denotes the value of the function $p_X$ at~$x$.
%Informally, we occasionally refer to the function $p_X$ as $p_X(x)$.
We use the notations $p_{X|Y}$ and $p_{X|Y}(x|y)$ equivalently.
% Equal to the first order in the exponent.
In addition, we shall follow the notation in~\cite{Cover06B} to specify that two sequences $a_k$ and $b_k$ are approximately equal in the exponent
if $\lim_{k \to \infty}\tfrac{1}{k} \log \tfrac{a_k}{b_k}=0$.
%Added to notation
As an example to illustrate this notation,
consider the sequences $a_k=2^{3k+\sqrt{k}}$ and $b_k=2^{3k}$,
and check that $\lim_{k \to \infty}\tfrac{1}{k} \log \tfrac{a_k}{b_k}=\lim_{k \to \infty}\tfrac{1}{\sqrt{k}}=0$,
what implies that they agree to first order in the exponent.
Further, throughout this work we shall denote the uniform distribution by $u$.
Last but not least, we shall use the notation $x^n$ to denote a sequence $x_1,x_2,\ldots,x_n$.

% Bayes Decision Theory
Having adopted these conventions, now we recall the basics on BDT. Namely, BDT is a statistical method that, fundamentally, uses a probabilistic
model to analyze the making of decisions under uncertainty and the costs associated with those decisions~\cite{Berger85B,Duda01B}. In general, Bayes
decision principles may be formulated in the following terms.
Consider the uncertainty refers to an~\emph{unknown} parameter modeled by an r.v.~$X$. In decision\hyph theoretic terminology, this is also
known as~\emph{state of nature}.
Let~$Y$ be another r.v.\ modeling an \emph{observation} or measurement on the state of nature.
Suppose that, given a particular observation~$y$, we are required to make a decision on the unknown.
Let~$\hat{x}$ denote the estimator of~$X$, that is, the rule that provides a decision or estimate~$\hat{x}(y)$ for every possible observation~$y$.
Clearly, any decision will be accompanied by a cost.
This is captured by the~\emph{loss function}~$d\colon(x,\hat{x})\mapsto d(x,\hat{x})$,
%This is captured by the~\emph{loss function}~$d(x,\hat{x}(y)),$
which measures how costly the decision~$\hat{x} = \hat{x}(y)$ will be when the unknown is~$x$.
However, since the actual loss incurred by a decision cannot be calculated with absolute certainty at the time the decision is made, BDT contemplates the
average loss associated with this decision.
% /Recursos/Berger1.pdf, page 9
Concretely, the \emph{Bayes conditional risk} for an estimator~$\hat{x}$ is defined in the discrete case as
$$\mathcal{R}(y) = \oE[d(X,\hat{x}(y))|y] = \sum_{x} p_{X|Y}(x|y)\,d(x,\hat{x}(y)),$$
where the expectation is taken over the \emph{posterior} probability distribution~$p_{X|Y}$. According to this, the \emph{Bayes risk} associated with that
estimator is defined as the average of the Bayes conditional risk over all possible observations~$y$, that is,
$$\mathcal{R} = \oE \oE[d(X,\hat{x}(Y))|Y] = \sum_{x,y} p_{X\,Y}(x,y)\,d(x,\hat{x}(y)),$$
where the expectation is additionally taken over the probability distribution of~$Y.$ Based on this definition, an estimator is called~\emph{Bayes estimator}
or \emph{Bayes decision rule}, if it minimizes the Bayes risk among all possible estimators.
It turns out that this optimal estimator is precisely
$$\hat{x}_\textnormal{Bayes}(y) = \argmin_{\hat{x}} \oE[d(X,\hat{x})|y],$$
for all~$y$; i.e., the Bayes estimator is the one that minimizes the Bayes conditional risk for every observation.

Once some of the basic elements in Bayes analysis have been examined, we would like to establish a connection between maximum a posteriori (MAP)
estimator and Bayes estimator. With this aim, first recall that a MAP estimator, as the name implies, is the estimator that maximizes the posterior distribution. Now
consider the loss function $d$ to be the Hamming distance between $x$ and $\hat{x}$,
which is an indicator function,
and recall that the expectation of an indicator r.v.\ is the probability of the event it is based on.
%DAVID 111027: Further emphasis on this
Mathematically,
$$\oE [d_\text{Hamming}(X,\hat{x})|y] = \oP \{X \neq \hat{x}|y\},$$
and consequently,
\begin{align}\label{eq:BayesMAP}
\hat{x}_\textnormal{MAP}(y) & =\argmin_{\hat{x}} \oP \{X \neq \hat{x}|y\}\nonumber\\
                                                              & = \argmax_{\hat{x}} \oP\{X=\hat{x}|y\}.
\end{align}
In conclusion, Bayes and MAP estimators coincide when the loss function is Hamming distance.

%%%%%%%%%%%%%%%%%%%%%%%%%%%%%%%%%%%%%%%%%%%%%%%%%%%%%%%%%%%%%%%%%%%%%%%%%%%%%%%%%%%%%%%%%%%%%%%%%%%%%%%%%%%%%%%%%%%%%%%%%%%%%%%%%%
%%%%%%%%%%%%%%%%%%%%%%%%%%%%%%%%%%%%%%%%%%%%%%%%%%%%%%%%%%%%%%%%%%%%%%%%%%%%%%%%%%%%%%%%%%%%%%%%%%%%%%%%%%%%%%%%%%%%%%%%%%%%%%%%%%
%4 MEASURING PRIVACY AS AN ATTACKER'S ESTIMATION ERROR
%%%%%%%%%%%%%%%%%%%%%%%%%%%%%%%%%%%%%%%%%%%%%%%%%%%%%%%%%%%%%%%%%%%%%%%%%%%%%%%%%%%%%%%%%%%%%%%%%%%%%%%%%%%%%%%%%%%%%%%%%%%%%%%%%%
%%%%%%%%%%%%%%%%%%%%%%%%%%%%%%%%%%%%%%%%%%%%%%%%%%%%%%%%%%%%%%%%%%%%%%%%%%%%%%%%%%%%%%%%%%%%%%%%%%%%%%%%%%%%%%%%%%%%%%%%%%%%%%%%%%
\section{Measuring Privacy as an Attacker's Estimation Error}\label{sec:Formulation}
\noindent
%Summary
This section presents our first contribution, a general framework that lays the foundation for the establishment of a unified measurement of privacy.
\begin{table}
\centering
\includegraphics[scale=0.70]{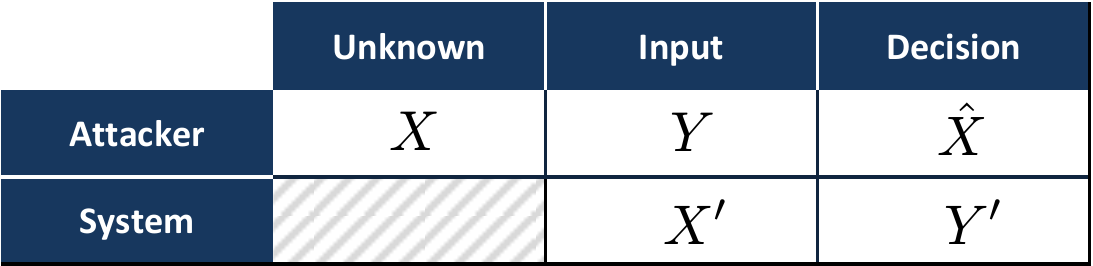}
\caption{Simplified representation of our notation.}\label{tab:notation:general}
\end{table}
However, it is not until Sec.~\ref{sec:Theory} where we shall show that a number of privacy criteria may be regarded as particular cases of our proposal.
Previously, Sec.~\ref{sec:Formulation:Notation} introduces our notation. Next, Sec.~\ref{sec:Formulation:Adversarial} describes the
adversarial model. In Sec.~\ref{sec:Formulation:Definition} we present our privacy metric, and finally, in Sec.~\ref{sec:Formulation:Example},
we illustrate the proposed formulation with a simple but insightful example.

%%%%%%%%%%%%%%%%%%%%%%%%%%%%%%%%%%%%%%%%%%%%%%%%%%%%%%%%%%%%%%%%%%%%%%%%%%%%%%%%%%%%%%%%%%%%%%%%%%%%%%%%%%%%%%%%%%%%%%%%%%%%%%%%%%%%%%%%%%%%%%%%%%%%%%%%%%%%%%%%%%%%%%%%%%%
\subsection{Mathematical Assumptions and Notation}\label{sec:Formulation:Notation}
%%%%%%%%%%%%%%%%%%%%%%%%%%%%%%%%%%%%%%%%%%%%%%%%%%%%%%%%%%%%%%%%%%%%%%%%%%%%%%%%%%%%%%%%%%%%%%%%%%%%%%%%%%%%%%%%%%%%%%%%%%%%%%%%%%%%%%%%%%%%%%%%%%%%%%%%%%%%%%%%%%%%%%%%%%%
\noindent
In this section, we provide the notation that we shall use throughout this work.
To this end, we first introduce the key actors of the proposed framework:

\begin{itemize}
\item a \emph{user}, who wishes to protect their privacy;
\item a (trusted) \emph{system}, to which each user entrusts their private data for its protection; %the unique purpose of this entity is to guarantee the privacy of the user,
the unique purpose of this entity is to guarantee the privacy of the user, and with this aim, the system may use any privacy\hyph preserving mechanism at its disposal;
\item and an \emph{attacker}, who strives to disclose private information about this user.
\end{itemize}

To clarify the elements involved in our framework, consider a conceptually\hyph simple approach to anonymous Web browsing,
consisting in a TTP acting as an intermediary between Internet users and Web servers.
From the perspective of our model, the users would be those subscribed to the anonymous proxy;
the system would be this proxy; and the attackers those servers that attempt to compromise users' privacy
from their Web browsing activity.

% 121002 Claudia : Remove the following paragraph.
%Having provided an example of said elements, next we shall introduce briefly the concepts of \emph{hard privacy} and \emph{soft privacy}~\cite{Mina10PHD}.
%The distinction between these two concepts will allow us to establish a connection between the actors \emph{user} and \emph{system} of our model.
%Specifically, hard privacy, which is also regarded as \emph{data minimization}, refers to the assumption that the user mistrusts any communicating entity and, consequently, endeavors to protect their privacy on their own.
%On the other hand, soft privacy assumes that the user entrusts their private data to an entity, which is thereafter responsible for the protection of their data.
%Put differently, and in terms of our framework, the concept of hard privacy boils down to the particular case in which the user adopts the role of the system; whereas the more general case in which the user does not coincide with the system corresponds to the assumptions of soft privacy.

%This is summarized in Table~\ref{tab:notation:general}.
In the following, the term r.v.\ is used with full generality to include categorical or numerical data, vectors, tuples or sequences of mixed components, but for mathematical simplicity we shall henceforth assume that all r.v.'s in the paper have finite alphabets.

\begin{itemize}

  \item The \emph{attacker's unknown} or \emph{uncertainty} is denoted by the r.v.\ $X$, which models the private information about a user that the attacker wishes to ascertain.

  \item The \emph{system's input} is represented by the r.v.\ $X'$ and refers to user's data required by the system to make a decision.

  \item The \emph{systems's decision} is modeled by the r.v.\ $Y'$ and denotes disclosed information, perhaps part of~$X'$, or a perturbation.

%  \item The \emph{attacker's input} or \emph{observation} is denoted by the r.v.~$Y$ and captures any evidence or measurement the attacker has about the unknown.
%  In some circumstances, this observation may be directly the information disclosed by the system, that is, $Y=Y'$.
%%  In other cases, the observation may be a perturbation of~$Y'$, perhaps together with background knowledge the attacker may have.
%Put another way, the only information available to the attacker is exactly that revealed by the system. In other cases, the attacker's input may be a perturbation of $Y'$, perhaps together with background knowledge the attacker may have acquired. In such cases, we have that $Y \neq Y'$.
  \item The \emph{attacker's input} is denoted by the r.v.\ $Y$ and captures any evidence or measurement the attacker has about the unknown.
  As its name indicates, this variable models the information that serves as input for the adversary to ascertain $X$.
  In some cases, $Y$ may be directly the information revealed by the system, i.e., $Y=Y'$.
  That is, the only information available to the attacker is exactly that disclosed by the system.
  In other circumstances, the attacker may observe a perturbed version of $Y'$,
  maybe together with background knowledge about the unknown.
  In such cases, we have $Y\neq Y'$.
  Since the attacker's input is, in fact, the information \emph{observed} by the attacker, directly from the system or indirectly from other sources,
  throughout this work we shall use the terms \emph{attacker's input} and \emph{attacker's observation} indistinguishably to refer to the variable $Y$.

  \item The \emph{attacker's decision} is modeled by the r.v.~$\hat{X}$ and represents the attacker's estimate of~$X$ from~$Y$.

\end{itemize}

\begin{table}
\centering
\includegraphics[width=1.00\columnwidth]{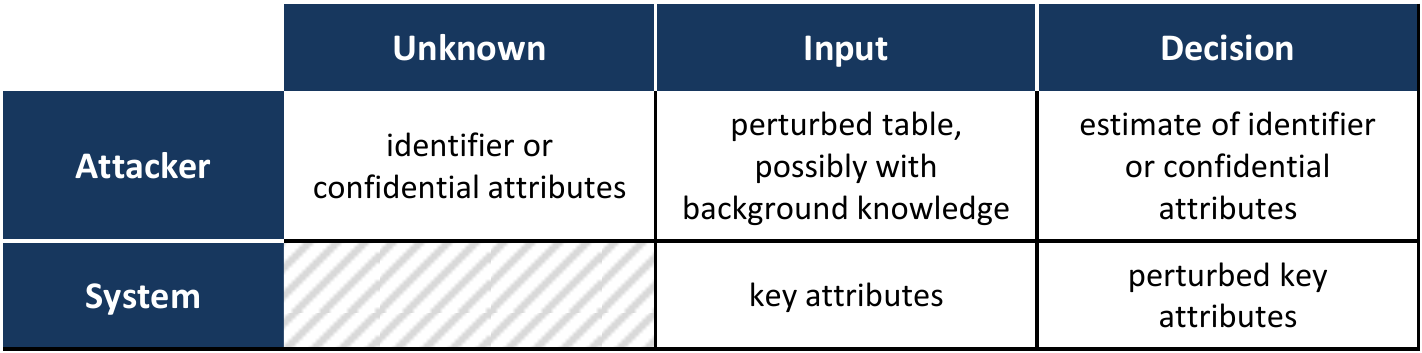}
\caption{Description of the variables used in our notation in the special case of SDC. Often, $X=X'$ and $Y=Y'$.}
\label{tab:notation:SDC}
\end{table}

% Example of SDC
In order to clarify this notation, we provide an example in which the above variables are put in the context of SDC.
In this scenario, the data publisher plays the role of the system.
Concretely, $X$~may represent identifying or confidential attribute values the attacker
endeavors to ascertain with regard to an individual appearing in a released table.
The individuals contained in this table are what we call users.
The system's input becomes now the key attribute values that the publisher has about the individuals.
On the other hand, $Y'$~is the perturbed version of those values, which jointly with the (unperturbed) confidential attribute
values, constitute the released table.
Furthermore, the attacker's input consists of the released table and, possibly, background knowledge the privacy attacker may have.
For example, this could be the case of a voter registration list.
In the end, the attacker's decision is the estimate of~$X$. All this information is shown in Table~\ref{tab:notation:SDC}.

\begin{figure}
\centering
\includegraphics[width=1.00\columnwidth]{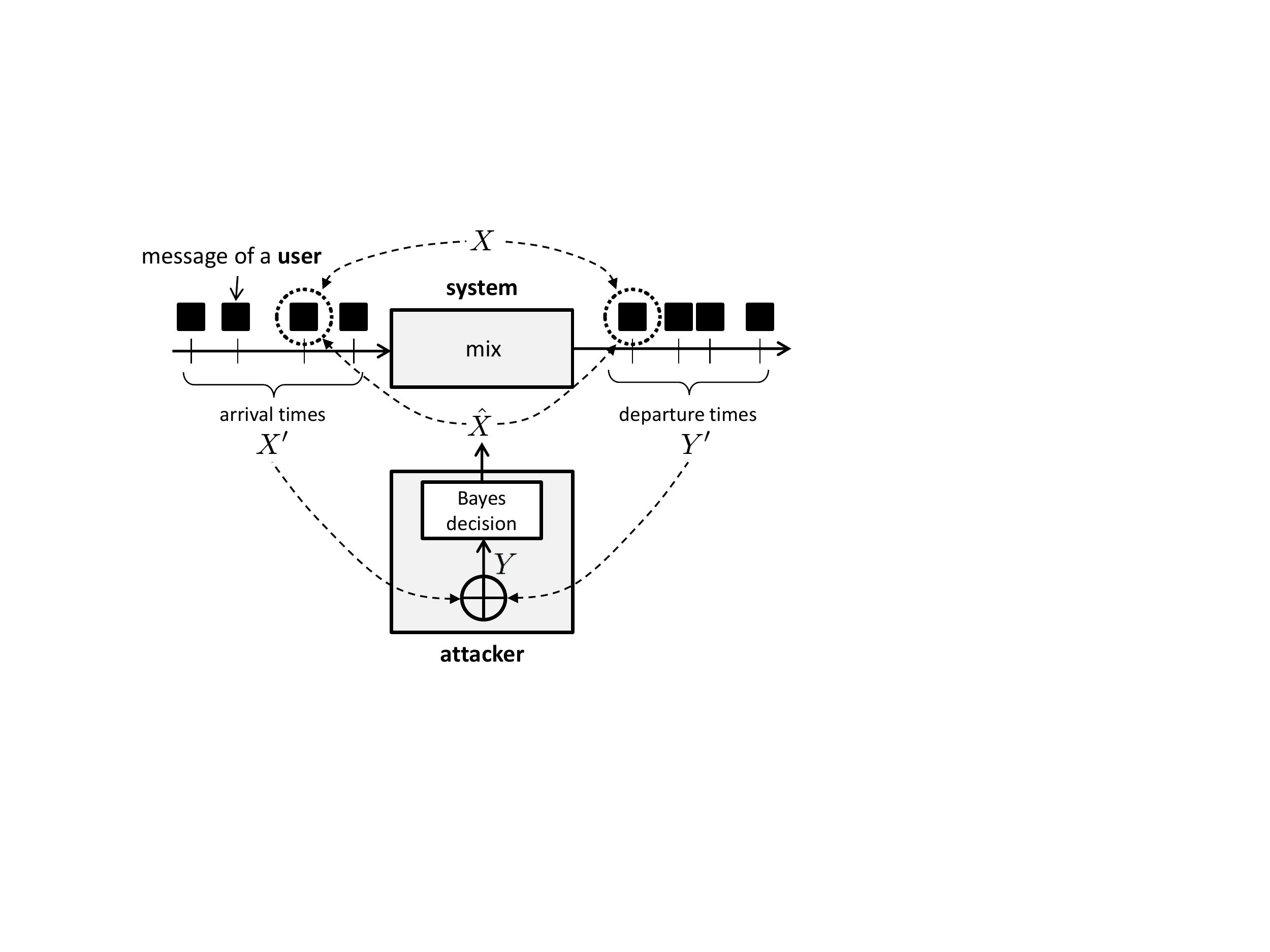}
\caption{Our framework is put in the context of mixes.}
\label{fig:mix_model}
\end{figure}

% Example of LBS
Similarly, now we specify the variables of our framework in the special case of a mix.
Under this scenario, the mix represents the system, whose objective is to hide the correspondence between the incoming and outgoing messages.
Precisely, the attacker's uncertainty is this correspondence.
The system's input and system's decision are the arrival and departure times of the messages, respectively.
On the other hand, the information available to the attacker, i.e., the attacker's observation $Y$, consists of $X'$, $Y'$ and the design parameters of the mix.
Finally, $\hat{X}$ is the attacker's decision on the correspondence between the messages.
This is depicted in Fig.~\ref{fig:mix_model} and summarized in~Table~\ref{tab:notation:mix}.
%%%%%%%%%%%%%%%%%%%%%%%%%%%%%%%%%%%%%%%%%%%%%%%%%%%%%%%%%%%%%%%%%%%%%%%%%%%%%%%%%%%%%%%%%%%%%%%%%%%%%%%%%%%%%%%%%%%%%%%%%%%%%%%%%%%%%%%%%%%%%%%%%%%%%%%%%%%%%%%%%%%%%%%%%%%
\subsection{Adversarial Model}\label{sec:Formulation:Adversarial}
%%%%%%%%%%%%%%%%%%%%%%%%%%%%%%%%%%%%%%%%%%%%%%%%%%%%%%%%%%%%%%%%%%%%%%%%%%%%%%%%%%%%%%%%%%%%%%%%%%%%%%%%%%%%%%%%%%%%%%%%%%%%%%%%%%%%%%%%%%%%%%%%%%%%%%%%%%%%%%%%%%%%%%%%%%%
\noindent
The consideration of a framework that encompasses a variety of privacy criteria necessarily requires the formalization of the attacker's model.
In this spirit, we now proceed to present the parameters that characterize this model.

Firstly, we shall contemplate an adversarial model in which the attacker uses a Bayes (best) decision rule.
Conceptually, this corresponds to the estimation made by an attacker that uses optimally the available information, as
we formally argued in Sec.~\ref{sec:Background:BDT}.
Namely, for every possible decision of the system resulting in an observation~$y$, the attacker will make a Bayes decision~$\hat{x}(y)$ on~$X$.
%Note that, from the point of view of the attacker, this estimator is the \emph{best} decision rule,
%since it minimizes our measure of privacy.
With regard to this attacker's decision rule, we would like to remark the fact that,
whereas it is a deterministic estimator, the system's decision is assumed to be a \emph{randomized} perturbation rule given by~$p_{Y^{'}|X^{'}}.$
As a consequence of this, it is clear that the system does not leak any private information when deciding $Y'$, provided
that $Y'$~and $X'$~are statistically independent.
%will provide ideal privacy if, and only if, $Y'$~and $X'$~are statistically independent.

Secondly, as explained in Sec.~\ref{sec:Background:BDT}, we shall require to evaluate the cost of each decision made by the attacker.
For this purpose, we consider the \emph{attacker's distortion function}~$d_\textnormal{A}\colon (x,\hat{x})\mapsto d_\textnormal{A}(x,\hat{x})$,
which measures the degree of dissatisfaction that the attacker experiences when~$X=x$ and~$\hat{X}=\hat{x}(y)$.
Similarly, we contemplate the \emph{system's distortion function}~$d_\textnormal{S} \colon (x',y') \mapsto d_\textnormal{S} (x',y')$,
which reflects the extent to which the system, and therefore the user, is discontent when~$Y'=y'$ and~$X'=x'$.

\begin{table}
\centering
\includegraphics[width=1.00\columnwidth]{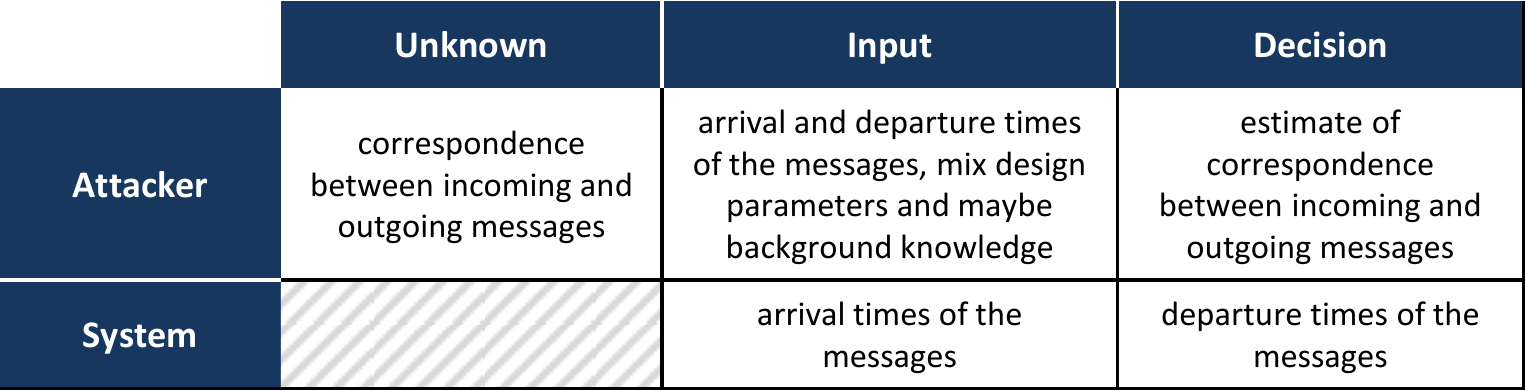}
\caption{Description of the variables used in our notation in the special case of mixes.}\label{tab:notation:mix}
\end{table}

%DAVID 111027: Moved from next section to this one, at Jordi's request
A crucial distinction in the type of attacker's distortion function $d_{\textnormal{A}}$ considered will be
whether it captures a sort of geometry over the symbols of the alphabet, or not.
%As shown in this figure, our first consideration relies upon an interesting property of the attacker's distortion function.
%In particular, we deemed reasonable to distinguish between those functions that capture a sort of geometry of the symbols of the alphabet, %and those which do not.
The most evident example of distortion function that does not take into account this geometry is the Hamming function, which we already
introduced at the end of Sec.~\ref{sec:Background:BDT}.
Concretely, this binary metric just indicates whether~$x$ and $\hat{x}$~coincide, and provides no more information about the discrepancy between them.
On the other hand, the squared error loss~$d_\textnormal{A}(x,\hat{x})=(x-\hat{x})^2$ and
the absolute error loss $d_\textnormal{A}(x,\hat{x})=|x-\hat{x}|$~are just two commonly\hyph used examples of distortion functions that do rely or induce a certain geometry.

%%%%%%%%%%%%%%%%%%%%%%%%%%%%%%%%%%%%%%%%%%%%%%%%%%%%%%%%%%%%%%%%%%%%%%%%%%%%%%%%%%%%%%%%%%%%%%%%%%%%%%%%%%%%%%%%%%%%%%%%%%%%%%%%%%%%%%%%%%%%%%%%%%%%%%%%%%%%%%%%%%%%%%%%%%%
\subsection{Definition of our Privacy Criterion}\label{sec:Formulation:Definition}
%%%%%%%%%%%%%%%%%%%%%%%%%%%%%%%%%%%%%%%%%%%%%%%%%%%%%%%%%%%%%%%%%%%%%%%%%%%%%%%%%%%%%%%%%%%%%%%%%%%%%%%%%%%%%%%%%%%%%%%%%%%%%%%%%%%%%%%%%%%%%%%%%%%%%%%%%%%%%%%%%%%%%%%%%%%
\noindent
Bearing in mind the above considerations, and consistently with Sec.~\ref{sec:Background:BDT}, we define
\emph{conditional privacy} as
\begin{equation}
\mathcal{P}(y) = \oE[d_\textnormal{A}(X,\hat{x}(y))|y],
\label{eq:conditional_privacy}
\end{equation}
which is the estimation error incurred by the attacker, conditioned on the observation~$y$. Based on this definition, we contemplate two possible measures of
privacy. In particular, we define \emph{worst\hyph case privacy} as
\begin{equation}
\mathcal{P}_\textnormal{min} = \min_y \mathcal{P}(y).
\label{eq:worst_privacy}
\end{equation}
On the other hand, we define \emph{average privacy} as
\begin{equation}
\mathcal{P}_\textnormal{avg} = \oE \mathcal{P}(Y) = \oE d_\textnormal{A} (X,\hat{x}(Y)),
\label{eq:average_privacy}
\end{equation}
which is the average of the conditional privacy over all possible observations~$y$.

In order to measure the utility loss caused by the perturbation of the original data, we define the \emph{average distortion} as
\begin{equation}
\mathcal{D} = \oE d_\textnormal{S} (X',Y').
\label{eq:average_distortion}
\end{equation}

According to these definitions, a privacy\hyph protecting system and an attacker would adopt the following strategies.
Namely, the system would select the decision rule~$p_{Y^{'}|X^{'}}$ that maximizes either the average privacy or the worst\hyph case privacy,
while not allowing the average distortion to exceed a certain threshold.
On the other hand, the attacker would choose the Bayes estimator, which would lead to the minimization of \emph{both} measures of privacy.
The reason behind this is that the Bayes estimator also minimizes the conditional privacy, as stated in Sec.~\ref{sec:Background:BDT}.

On a different note, we would like to remark that a privacy risk~$\mathcal{R}$ in lieu
of~$\mathcal{P}$ could be defined for~$-d_\textnormal{A} (x,\hat{x}(y))$ instead of~$d_\textnormal{A} (x,\hat{x}(y))$. An analogous argument justifies the use of utility instead of distortion.

Last but not least, we would also like to note that, in the special case when the unknown variable~$X$ models the identity of a user, our measure of privacy may be regarded, in fact, as a measure of anonymity.
%%%%%%%%%%%%%%%%%%%%%%%%%%%%%%%%%%%%%%%%%%%%%%%%%%%%%%%%%%%%%%%%%%%%%%%%%%%%%%%%%%%%%%%%%%%%%%%%%%%%%%%%%%%%%%%%%%%%%%%%%%%%%%%%%%%%%%%%%%%%%%%%%%%%%%%%%%%%%%%%%%%%%%%%%%%
\subsection{Example}\label{sec:Formulation:Example}
%%%%%%%%%%%%%%%%%%%%%%%%%%%%%%%%%%%%%%%%%%%%%%%%%%%%%%%%%%%%%%%%%%%%%%%%%%%%%%%%%%%%%%%%%%%%%%%%%%%%%%%%%%%%%%%%%%%%%%%%%%%%%%%%%%%%%%%%%%%%%%%%%%%%%%%%%%%%%%%%%%%%%%%%%%%
\noindent
Next, we present a simple example that sheds some light on the formulation introduced in the previous sections.

For the sake of simplicity, consider~$X'=X$, that is, the system's input is the confidential information that needs to be protected. Suppose that~$X$ is a
binary r.v.\ with $\oP\{X=0\}=\oP\{X=1\}=\nicefrac{1}{2}$. In order to hinder
privacy attackers in their efforts to ascertain~$X$, for each possible outcome~$x$, the system will disclose a perturbed version~$y'$.
Namely, with probability~$p$ the system will decide to reveal the complementary value of~$x$,
whereas with probability~$1-p$ no perturbation will be applied, i.e., $y'=x$.
Note that, in this example, the system's decision rule is completely determined by~$p$, for which we conveniently impose the condition~$0 \leqslant p<\nicefrac{1}{2}$.

At this point, we shall assume that the attacker only has access to the disclosed information~$Y'$,
and therefore the attacker's input $Y$ boils down to it.
We anticipate that, throughout this work, this supposition will be usual.
In addition, we shall consider the attacker's distortion function to be the Hamming distance.
However, as commented on in Sec.~\ref{sec:Background:BDT}, this implies that the Bayes estimator matches the MAP estimator.
According to this observation, it is easy to demonstrate that the attacker's best decision is~$\hat{X}=Y.$
Therefore, the average privacy~(\ref{eq:average_privacy}) becomes
$$\mathcal{P}_\textnormal{avg}=\oP \{X\neq \hat{X}\} = \oP \{X\neq Y\}= \oP \{X\neq Y'\}= p.$$
On the other hand, if we suppose that the system's distortion function is also the Hamming distance, from~(\ref{eq:average_distortion}), it follows that $$\mathcal{D}=\oP\{X'\neq Y'\} = \oP\{X\neq Y'\} = p.$$

\begin{figure}
\centering
\includegraphics[scale=0.50]{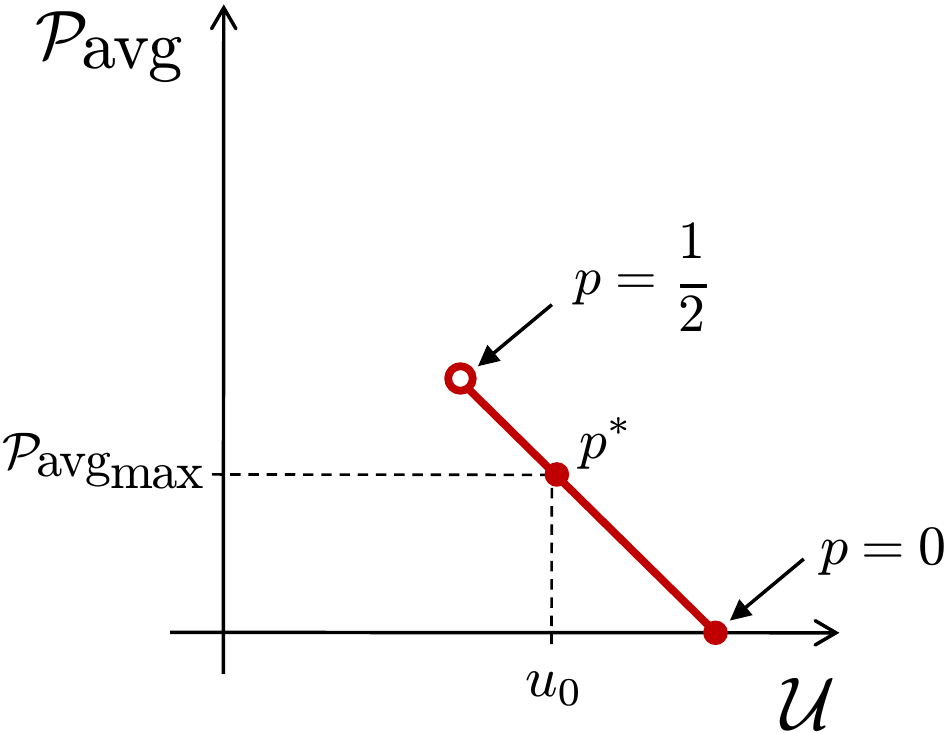}
\caption{Representation of the trade\hyph off curve between privacy and utility for the example.}\label{fig_example_tradeoff}
\end{figure}

Based on these two results, we now proceed to describe the strategy that the attacker would follow. To this end, we define the \emph{average utility}~$\mathcal{U}$ as $1-\mathcal{D}$.
According to this, the system would strive to maximize the average privacy with respect to~$p$, subject to the constraint~$\mathcal{U} \geqslant u_0$.
Fig.~\ref{fig_example_tradeoff} illustrates this simple optimization problem by showing the trade\hyph off curve between privacy and utility.
In this example, it is straightforward to verify that the optimal value of
average privacy is~$\mathcal{P}_{\textnormal{avg}_\textnormal{max}}=1-u_0,$ for~$\nicefrac{1}{2}< u_0 \leqslant 1.$

%%%%%%%%%%%%%%%%%%%%%%%%%%%%%%%%%%%%%%%%%%%%%%%%%%%%%%%%%%%%%%%%%%%%%%%%%%%%%%%%%%%%%%%%%%%%%%%%%%%%%%%%%%%%%%%%%%%%%%%%%%%%%%%%%%
%%%%%%%%%%%%%%%%%%%%%%%%%%%%%%%%%%%%%%%%%%%%%%%%%%%%%%%%%%%%%%%%%%%%%%%%%%%%%%%%%%%%%%%%%%%%%%%%%%%%%%%%%%%%%%%%%%%%%%%%%%%%%%%%%%
%5 Theory
%%%%%%%%%%%%%%%%%%%%%%%%%%%%%%%%%%%%%%%%%%%%%%%%%%%%%%%%%%%%%%%%%%%%%%%%%%%%%%%%%%%%%%%%%%%%%%%%%%%%%%%%%%%%%%%%%%%%%%%%%%%%%%%%%%
%%%%%%%%%%%%%%%%%%%%%%%%%%%%%%%%%%%%%%%%%%%%%%%%%%%%%%%%%%%%%%%%%%%%%%%%%%%%%%%%%%%%%%%%%%%%%%%%%%%%%%%%%%%%%%%%%%%%%%%%%%%%%%%%%%
\section{Theoretical Analysis}\label{sec:Theory}
\noindent
In this section, we present our second contribution, namely, the interpretation of several well\hyph known privacy criteria as particular cases of our more general
definition of privacy.
The arguments behind the justification of these privacy metrics as a particularization of our criterion are based on numerous concepts from the fields of
information theory, probability theory and BDT.
In this section, we therefore approach this issue from a theoretical perspective;
however, we refer those readers not particularly interested in the mathematical details to Sec.~\ref{sec:Guide}.

\begin{figure}
\centering
\includegraphics[scale=0.48]{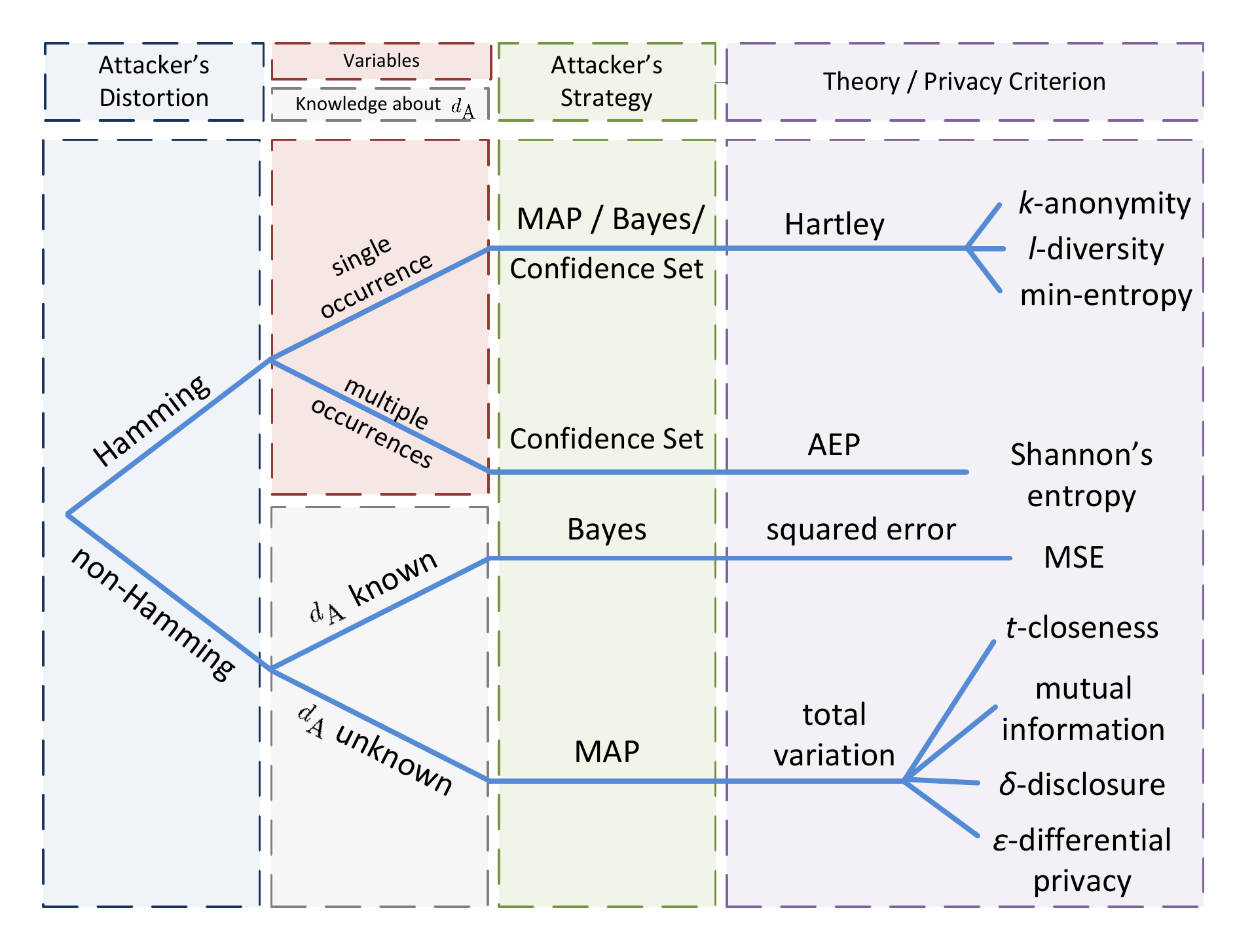}
\caption{The arguments that lead to the interpretation of several privacy metrics as particular cases of our definition of privacy are conceptually organized in the above points.
As can be observed, these arguments clearly depend on the attacker's distortion function,
namely on the geometry of this function (Hamming or non\hyph Hamming)
and on the knowledge the user has about it, i.e., it is known or unknown to the user.
Other parameters include the nature of the variables of our framework, and obviously the attacker's strategy.}\label{fig:sketch}
\end{figure}

For a comprehensive exposition of these arguments, the underlying assumptions and concepts will be expounded in a systematic manner, following the points sketched in Fig.~\ref{fig:sketch}.
As mentioned in Sec.~\ref{sec:Formulation:Adversarial} and illustrated by the first branch of the tree depicted in this figure,
our starting point makes the significant distinction between attacker's distortion measures based on the Hamming distance and the rest,
according to whether we wish to capture a certain, gradual measure of distance between alphabet values beyond sheer symbol equality.
It is important to recall from Sec.~\ref{sec:Background:BDT} that
in the case of a Hamming distortion measure, expected distortion boils down to probability of error,
yielding a different class of estimation problems.

Bearing in mind the above remark, in Sec.~\ref{sec:Theory:Hamming} we shall contemplate the case when the attacker's distortion function is the Hamming distance, whereas in Sec.~\ref{sec:Theory:noHamming} we shall deal with the more general case in which~$d_{\textnormal{A}}$ can be any other distortion function.
In the special case of Hamming distance, we consider two alternatives for the variables in Table~\ref{tab:notation:general}:
single\hyph occurrence and multiple\hyph occurrence data.
The former case considers the variables to be tuples of a small number of components, and the latter case assumes that these variables are sequences of data.
In the scenario of single\hyph occurrence data, we shall establish a connection between Hartley's entropy and our privacy metric,
which will allow us to interpret $k$\hyph anonymity, $l$\hyph diversity and min\hyph entropy criteria as particular cases of our framework.
The arguments that will enable us to justify this connection stem from MAP estimation, BDT and the concept of confidence set.
On the other hand, when we consider multiple\hyph occurrence data,
we shall use the asymptotic equipartition property (AEP) to argue that
the Shannon entropy, as a measure of privacy, is a characterization of the cardinality of a high\hyph confidence set of sequences.

In the more general case in which the attacker's distortion function is not the Hamming distance, we shall explore two possible scenarios.
On the one hand, we shall consider the case where this function is known to the system.
Under the assumption of a Bayes attacker's strategy, we shall use BDT to justify the system's best decision rule.
On the other hand, we shall contemplate the case in which the attacker's distortion function is unknown to the system.
Specifically, this scenario will allow us to connect our framework to several privacy criteria
through the concept of total variation, provided that the attacker uses MAP estimation.
\begin{figure*}
\centering
\includegraphics[scale=0.65]{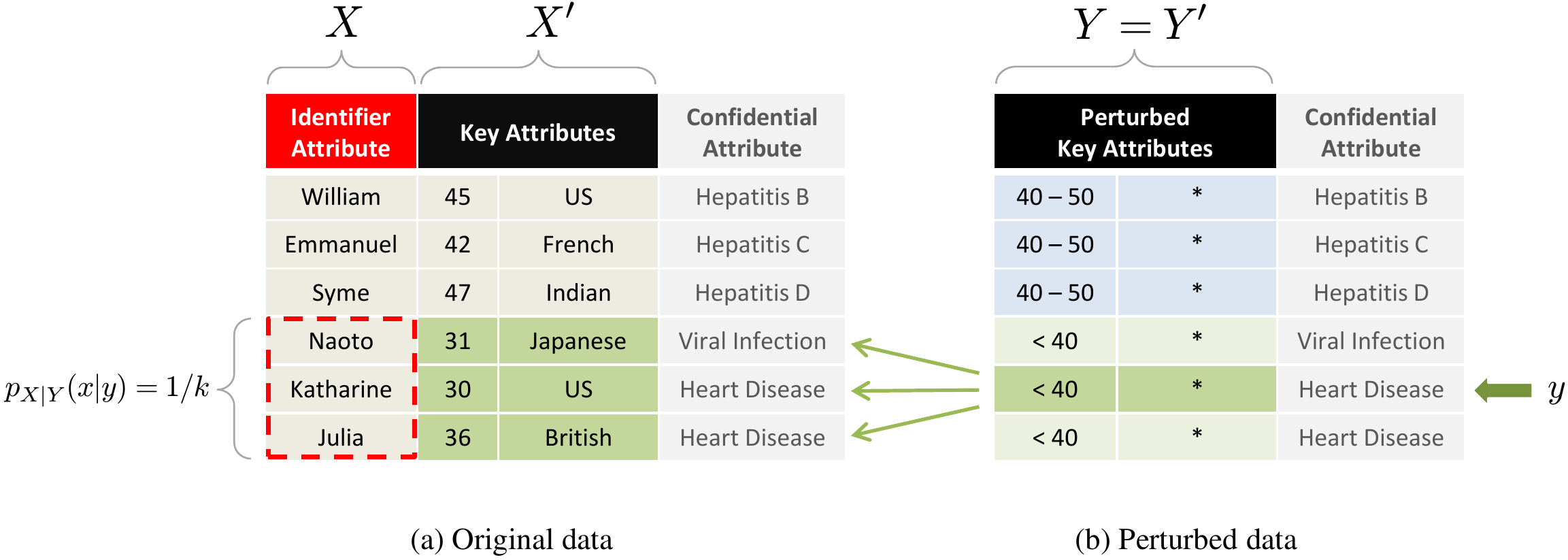}
\caption{A data publisher plans to release a 3\hyph anonymized microdata set. To this end, the publisher must enforce that, for a given tuple of key attribute values
in~(b), the probability of ascertain the identifier value of the corresponding record
in~(a) must be at most~\nicefrac{1}{3}.}
\label{fig:k-anonymity}
\end{figure*}
%%%%%%%%%%%%%%%%%%%%%%%%%%%%%%%%%%%%%%%%%%%%%%%%%%%%%%%%%%%%%%%%%%%%%%%%%%%%%%%%%%%%%%%%%%%%%%%%%%%%%%%%%%%%%%%%%%%%%%%%%%%%%%%%%%
%4.1 Hamming Distortion
%%%%%%%%%%%%%%%%%%%%%%%%%%%%%%%%%%%%%%%%%%%%%%%%%%%%%%%%%%%%%%%%%%%%%%%%%%%%%%%%%%%%%%%%%%%%%%%%%%%%%%%%%%%%%%%%%%%%%%%%%%%%%%%%%%
\subsection{Hamming Distortion}\label{sec:Theory:Hamming}
\noindent
In this section, we shall analyze the special case when the attacker's distortion function is the Hamming distance,
commented on in Secs.\ \ref{sec:Background:BDT} and~\ref{sec:Formulation:Adversarial}.
In addition, we shall contemplate two cases for the variables of our framework:
single\hyph occurrence and multiple\hyph occurrence data.

\SpaceBeforeSection

%%%%%%%%%%%%%%%%%%%%%%%%%%%%%%%%%%%%%%%%%%%%%%%%%%%%%%%%%%%%%%%%%%%%%%%%%%%%%%%%%%%%%%%%%%%%%%%%%%%%%%%%%%%%%%%%%%%%%%%%%%%%%%%%%%
%4.1.1 Single Occurrence
%%%%%%%%%%%%%%%%%%%%%%%%%%%%%%%%%%%%%%%%%%%%%%%%%%%%%%%%%%%%%%%%%%%%%%%%%%%%%%%%%%%%%%%%%%%%%%%%%%%%%%%%%%%%%%%%%%%%%%%%%%%%%%%%%%
\subsubsection{Single Occurrence}\label{sec:Theory:Hamming:Single}
\noindent
This section considers the scenario in which the variables defined in Sec.~\ref{sec:Formulation:Notation} are tuples of a relatively small number of components, including both categorical and numerical data, defined on a finite alphabet.
In order to establish a connection between some of the most popular privacy metrics and our criterion,
first we shall introduce the concept of confidence set and briefly recall a riveting generalization of Shannon's entropy.

% Confidence set
% Once we model the attacker's target information X as a (possibly multiletter) r.v., a conceivable measure of privacy consists in the cardinality of a confidence set.
Consider an r.v.~$X$ taking on values in the alphabet $\mathcal{X}$.
A \emph{confidence set} $\sC$ with confidence $p$ is defined as a subset of $\mathcal{X}$ such that $\oP\{X \in \sC\}=p$.
In the case of continuous\hyph valued random scalars, confidence sets commonly take the form of intervals.
%For discrete r.v.'s, the smallest sets will be obtained simply by choosing the most likely members.
In these terms, it is clear that a privacy attacker aimed at ascertaining $X$ will benefit the most from those confidence sets whose cardinality is reduced substantially with respect to the original alphabet size, with high confidence.
To connect the concept of confidence set to our interpretation of privacy as an attacker's estimation error,
consider an attacker model where the attacker only takes into account the shape of the PMF of the unknown $X$ to identify a confidence set $\sC$ for some desired confidence~$p$, and beyond that, assumes all the included members equally relevant.
This last assumption may be interpreted as an investigation on a tractable list of potential identities, carried out in parallel.
MAP estimation within that set, considering it uniformly distributed, leads to an estimation error of
$1-\tfrac{1}{|\sC|}$, that is, a bijection of its cardinality.

% R\'enyi's entropy
In our interpretations, we further use the R\'enyi entropy, a family of functionals widely used in information theory as a measure of uncertainty.
More specifically, R\'enyi's entropy of order $\alpha$ is defined as
\begin{equation*}
\oH_\alpha (X) = \frac{1}{1-\alpha} \log \sum_{i=1}^{n} p_X(x_i)^{\alpha},
\end{equation*}
where $p_X$ is the PMF of an r.v.\ $X$ that takes on values in the alphabet~$\mathcal{X}=\{x_1,\ldots,x_n\}$.
In the important case when $\alpha$ is~0, R\'enyi's entropy is essentially given by the support set of~$p_X$,
that is,
$$\oH_0(X)=\log \left|\{x\in \mathcal{X}: p_X (x)>0\}\right|.$$
In this particular case, R\'enyi's entropy is referred to as Hartley's entropy.
Evidently, when $p_X$~is strictly positive, the support set becomes the alphabet and $\oH_0(X)=\log n$.
Under this assumption, the Hartley entropy can be understood as a confidence set with $p=100\%$.
On the other hand, in the limit when $\alpha$ approaches~$1$, R\'enyi's entropy reduces to Shannon's
$$\oH_1(X)=-\sum_i p_X(x_i)\,\log p_X(x_i).$$
Lastly, in the limit as $\alpha$ goes to $\infty$, the R\'enyi entropy approaches the \emph{min\hyph entropy}
$$\oH_\infty(X)=\min_i -\log p_X(x_i)= -\log \max_i p_X(x_i).$$

We shall shortly interpret min\hyph entropy, Shannon's entropy and Hartley's entropy within our general framework of privacy
as an attacker estimation error, when Hamming distance is used as a distortion measure,
first for single occurrences of a target information, and later for multiple occurrences.
For now, we could loosely consider an attacker striving to ascertain the outcome of the finite\hyph alphabet r.v.~$X$,
and the effect of the dispersion of its PMF on such task.
Conceptually, we could then regard these three types of entropies simply as worst\hyph case, average\hyph case and best\hyph case
measurements of privacy, respectively, on account of the fact that
\begin{equation}\label{eqn:Theory:EntropyInequalities}
\oH_\infty(X) \leqslant \oH_1(X) \leqslant \oH_0(X),
\end{equation}
with equality if, and only if, $X$ is uniformly distributed.
More specifically, the min\hyph entropy $\oH_\infty(X)$ is the minimum of the \emph{surprisal} or \emph{self\hyph information}
 $-\log p_X(x_i)$,
whereas the Shannon entropy $\oH_1(X)$ is a weighted average of such logarithms, and finally,
the Hartley entropy $\oH_0(X)$ optimistically measures the cardinality of the entire set of possible values of $X$
regardless of their likelihood.

After showing the Hartley, Shannon and min entropies are particular cases of R\'enyi's entropy,
now we go on to describe a scenario that will allow us to relate our privacy
metric to an extensively\hyph used criterion.
Specifically, we focus on the important case of SDC, where the data publisher plays the system's role.
In this scenario, a data publisher wishes to release a microdata set and, before distributing it, the publisher applies some algorithm~\cite{Truta06PDM,Machanavajjhala06ICDE,Li07ICDE,Brickell08KDD,Dwork06A,Rebollo10KDE} to enforce the $k$\hyph anonymity requirement~\cite{Sweeney02UFKBS,Samarati01KDE}.
As mentioned in Sec.~\ref{sec:Background:SDC}, the objective of a linking attack is to unveil the identity of the individuals appearing in a released table by linking the records in this table to any public data set including identifiers.
Since $k$\hyph anonymity is aimed at protecting the data against this attack, in our scenario the attacker's unknown~$X$ becomes the user identity.
The other variables shown in Table~\ref{tab:notation:SDC} are as follows: $X'$~are the key attribute values, $Y'$~are the perturbed key attribute values, the attacker's observation~$Y$ is assumed to be~$Y'$, and finally, $\hat{X}$~is an estimate of the identity of a user.
Fig.~\ref{fig:k-anonymity} illustrates this particular case.

In order to protect the data set from identity disclosure, the algorithm must ensure that, for any observation~$y$ consisting in a tuple of perturbed key attribute values
in the released table, the identifier value of the corresponding record in the original table cannot be ascertained beyond
a subgroup of at least $k$~records.
As we shall see next, this requirement will be reflected mathematically by assuming that the probability distribution~$p_{X|Y}(\cdot|y)$ of the identifier value, conditioned on the observation~$y$, is the uniform distribution on a set of at least $k$~individuals.
% On a different note, we would like to remark that, if this requirement is met, the data are $k$\hyph anonymous regardless of any publicly available table with identifiers the attacker may have.
% 121002 Claudia: I would remove this claim...  What if the the adversary has a table in which together with the identifiers there is info that (co-)relates to the sensitive attributes? Even if you consider that the adversary does not learn additional sensitive info on some users with the released table (because he already knew that info for those users), the released table would help him gain information on _other_ users he did not know about. Simples case: I know the sensitive attributes for k-1 people, with the released table I learn additional info about a k-th person that has been put together with those k-1.
%For this reason, we may consider the more general case in which $Y$~consists of~$Y'$ and any background knowledge.
Lastly, we consider the more general case in which $Y$~consists of~$Y'$ and any background knowledge.

That said, our adversarial model contemplates an attacker who uses a MAP estimator, which, as shown in Sec.~\ref{sec:Background:BDT}, is equivalent to the Bayes estimator.
Under this model, given an observation~$y$, the conditional privacy~(\ref{eq:conditional_privacy}) becomes
\begin{equation}\label{eq:condprivacy1}
\mathcal{P}(y)=\oP\{X \neq \hat{x}(y)|y\}=1 - \max_x p_{X|Y}(x|y),
\end{equation}
which precisely is the MAP error~$\varepsilon_{_\textnormal{MAP}}$, conditioned on that observation~$y$;
in terms of min\hyph entropy, we may recast our metric as
$$\mathcal{P}(y)=\varepsilon_{_\textnormal{MAP}}=1-2^{-\oH_\infty(X|y)},$$
which shows that the concept of min\hyph entropy is intimately related to MAP decoding.
If we finally apply the aforementioned uniformity condition of~$p_{X|Y}(\cdot|y)$,
and assume that this PMF is the uniform distribution on a group of exactly $k$ individuals, that is, $u_i=1/k$~for all $i=1,\ldots,k$, then
\begin{equation*}
\mathcal{P}(y)=1-1/k=1-2^{-\oH_0(X|y)},
\end{equation*}
which expresses the conditional privacy in terms of Hartley's entropy.
In a nutshell, the $k$\hyph anonymity criterion may be interpreted as a special case of our privacy measure,
determined by this R\'enyi's entropy.

%%%%%%%%%%%%%%%%%%%%%%%%%%%%%%%%%%%%%%%%%%%%%%%%%%%%%%%%%%%%%%%%%%%%%%%%%%%%%%%%%%%%%%%%%%%%%%%%%%%%%%%%%%%%%%%%%%%%%%%%%%%%%%%%%%%%%%%%%%%%%%%%%%%%%%%%%%%%%%%%%%%%%%%
%l-diversity
%%%%%%%%%%%%%%%%%%%%%%%%%%%%%%%%%%%%%%%%%%%%%%%%%%%%%%%%%%%%%%%%%%%%%%%%%%%%%%%%%%%%%%%%%%%%%%%%%%%%%%%%%%%%%%%%%%%%%%%%%%%%%%%%%%%%%%%%%%%%%%%%%%%%%%%%%%%%%%%%%%%%%%%

After examining this first interpretation, next we shall explore an enhancement of $k$\hyph anonymity.
As argued in Sec.~\ref{sec:Background:BDT}, this criterion does not protect against confidential attribute disclosure.
In an effort to address this limitation, several privacy metrics were proposed.
In the remainder of this section, we shall focus on one of these approaches.
In particular, we shall consider the $l$\hyph diversity metric~\cite{Machanavajjhala06ICDE}, which builds on the $k$\hyph anonymity principle and aims at overcoming the attribute disclosure problem.
\begin{figure*}
\centering
\includegraphics[scale=0.65]{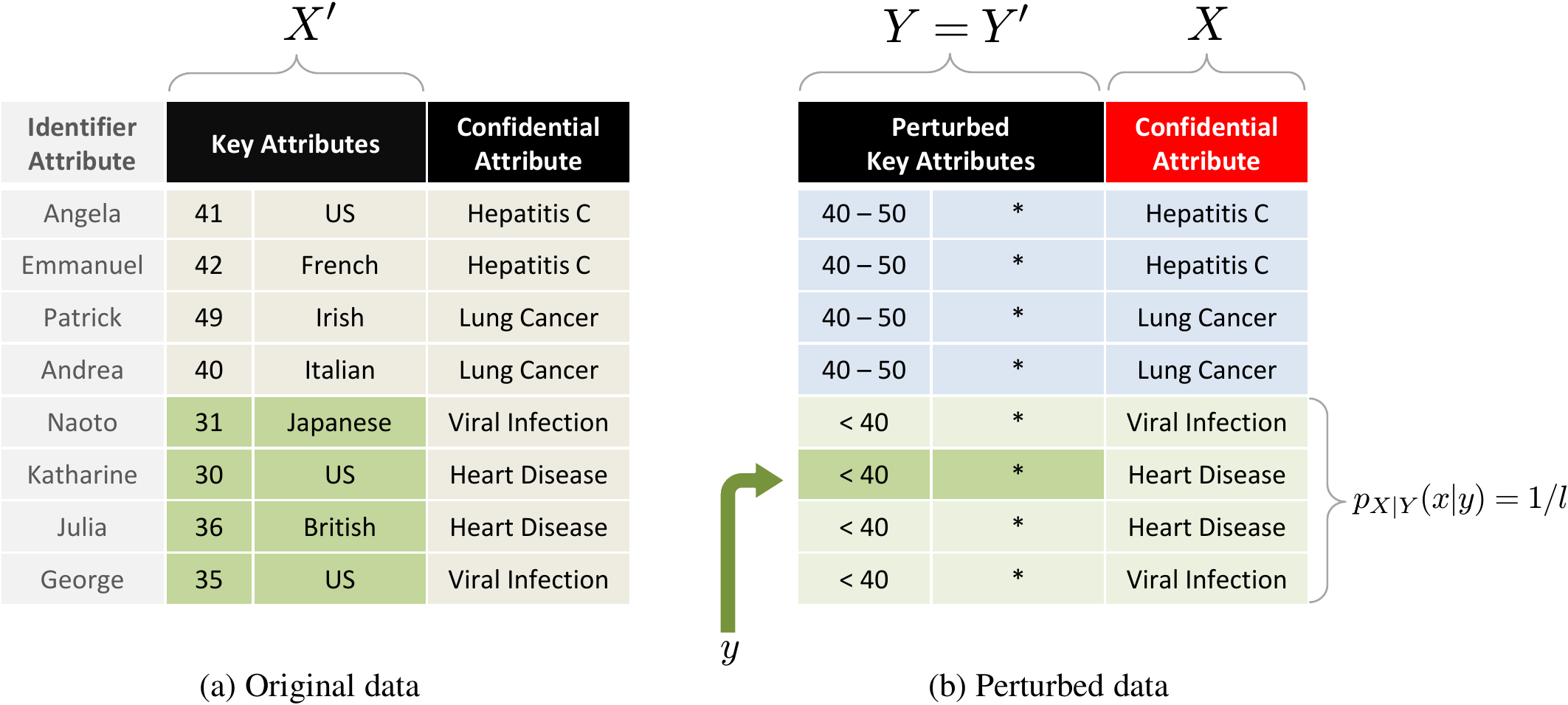}
\caption{In this example, the 2\hyph diversity principle is applied to a microdata set. In order to meet this requirement, we assume that, for each group of records with the same tuple of perturbed key attribute values, the probability distribution of the confidential attribute value in~(b) is the uniform distribution on a set of at least $2$~values.}
\label{fig:l-diversity}
\end{figure*}

%What's l-diversity and some of its different instantiations: l-distinct and entropy l-diversity
As mentioned in Sec.~\ref{sec:Background:SDC}, a microdata set satisfies $l$\hyph diversity~if, for each group of records
sharing a tuple of key attribute values in the perturbed table, there are at least $l$~``well\hyph represented'' values for each confidential attribute.
Depending on the definition of well\hyph represented, this criterion can reduce to distinct $l$\hyph diversity,
which is equivalent to $l$\hyph sensitive $k$\hyph anonymity, or be more restrictive.
Concretely, a microdata is said to meet the entropy $l$\hyph diversity requirement if, for each group of records with the same tuple of perturbed key attribute values,
the entropy of each confidential attribute is at least~$\log l$.

% Notation
In our new scenario, a data publisher, still playing the system's role, applies an algorithm on the microdata set to enforce the $l$\hyph diversity principle.
Since the aim of this criterion is to protect the data against attribute disclosure, we consider that the attacker's unknown~$X$ refers to the confidential attribute.
The other variables remain the same as in our previous interpretation.

% Assumptions
Having said that, we shall make the assumption that the $l$\hyph diversity requirement is met by enforcing that, for a given tuple~$y$
of perturbed key attribute values, the probability distribution~$p_{X|Y}(\cdot|y)$ of the confidential attribute within the group of records sharing this tuple is the uniform distribution
on a set of at least $l$~values.
This is depicted in Fig.~\ref{fig:l-diversity}.
Note that this assumption entails that the data fulfill both the distinct and entropy $l$\hyph diversity requirements.
Lastly, we shall suppose again that the attacker uses MAP estimator.

% Privacy derivations
As mentioned before, under the premise of a MAP attacker, our measure of conditional privacy boils down to the MAP error~\eqref{eq:condprivacy1}.
If we also apply the assumption above about the uniformity of~$p_{X|Y}(\cdot|y)$, and
suppose that this distribution is uniform on a group of $l$~individuals, then the conditional privacy yields
$$\mathcal{P}(y)=1-1/l=1-2^{-\oH_0(X|y)},$$
which expresses our privacy metric again in terms of Hartley's entropy.
In short, the $l$\hyph diversity criterion lends itself to be interpreted as a particular case of our more general privacy measure.
%characterized by Hartley's entropy, as in the case of $k$\hyph anonymity.

%Bearing in mind these two considerations, and given the observation~$y$, the conditional privacy reduces to the MAP error~\eqref{eq:condprivacy1}.
%Analogously to the justification of $k$\hyph anonymity, if we suppose further that the $l$\hyph diversity criterion is fulfilled with equality for
%the observation~$y$, that is, $p_{X|Y}(x|y)$~is the uniform distribution~$u_i=1/l$~for all $i=1,\ldots,l$, then
%$$\mathcal{P}(y)=1-1/l=1-2^{-\oH_0(X|y)},$$
%which again coincides with the $\varepsilon_{_\textnormal{MAP}}$ conditioned on the observation, and defines the conditional privacy in terms of Hartley's entropy.
%In short, the $l$\hyph diversity criterion lends itself to be interpreted as a particular case of our more general privacy measure.
%Lastly, we would like to stress that the concept of min\hyph entropy is tightly related to MAP decoding.
%Still under the assumption of a MAP attacker, when considering an arbitrary PMF $p_{X|Y}(x|y)$, not necessarily uniform,
%the conditional privacy is
%$$\mathcal{P}(y)=1-2^{-\oH_\infty(X|y)}.$$

\SpaceBeforeSection

%%%%%%%%%%%%%%%%%%%%%%%%%%%%%%%%%%%%%%%%%%%%%%%%%%%%%%%%%%%%%%%%%%%%%%%%%%%%%%%%%%%%%%%%%%%%%%%%%%%%%%%%%%%%%%%%%%%%%%%%%%%%%%%%%%
%4.1.2 Multiple Occurrence
%%%%%%%%%%%%%%%%%%%%%%%%%%%%%%%%%%%%%%%%%%%%%%%%%%%%%%%%%%%%%%%%%%%%%%%%%%%%%%%%%%%%%%%%%%%%%%%%%%%%%%%%%%%%%%%%%%%%%%%%%%%%%%%%%%
\subsubsection{Multiple Occurrences}\label{sec:Theory:Hamming:Multiple}
\noindent
%Brief outline of the subsection
In this section, we shall consider the case when the variables shown in Table~\ref{tab:notation:general} are sequences of categorical and numerical data but in a finite alphabet.
Recall from Sec.~\ref{sec:Background:BDT} that we use the notation~$X^k$ to denote a sequence $X_1,\ldots, X_k$.

% Motivating example of sequences of r.v.'s
The special case that we contemplate now could perfectly model the scenario in which a user interacts with an LBS provider,
through an intermediate system protecting the user's location privacy.
In this scenario, a user would submit queries along with their locations to the trusted system.
An example would be the query ``Where is the nearest parking garage?'', accompanied by the geographic coordinates of the user's current location.
As many approaches suggest in the literature of private LBSs, the system would perturb the user coordinates and submit them to the LBS provider.
Concordantly, we may choose Euclidean distance as the natural attacker's distortion measure.
Alternatively, if the attacker's interest lies in whether the user is at home, at work, shopping for groceries or at the movies, in order to profile their behavior,
or more simply, whether the user is at a given sensitive location or not, then the appropriate model for the location space becomes discrete, and Hamming distance is more suited.

In this context, the consideration of sequences of discrete r.v.'s in our notation makes sense.
Specifically, an attacker would endeavor to ascertain the sequence~$X^k$ of~$k$ unknown locations visited by the user, from the sequence~$Y'^k$ of~$k$ perturbed locations that the system would submit to the LBS.
Put differently, the attacker's unknown would be the location data the user conveys to the system, i.e., $X^k=X'^k$,
and the information available to the adversary the perturbed version of this data, that is, $Y^k=Y'^k$.

% Justification - Start by setting our notation
Having motivated the case of sequences of data, in this section we shall establish a connection between our metric and Shannon's entropy as a measure of privacy.
But in order to emphasize this connection, first we briefly recall one of the pillars of information theory: the asymptotic equipartition property~\cite{Cover06B},
which derives from the weak law of large numbers and results in important consequences in this field.

% AEP - The typical set
Consider a sequence $X^k$ of $k$ independent, identically distributed (i.i.d.) r.v.'s, drawn according to~$p_X$, with alphabet size~$n$.
Loosely speaking, the AEP states that among all possible $n^k$ sequences,
there exists a \emph{typical subset}~$\mathscr{T}_{\epsilon}^{k}$ of sequences almost certain to occur.
More precisely, for any $\epsilon>0$, there exists a~$k$ sufficiently large such that $\oP\{\mathscr{T}_{\epsilon}^{k}\} > 1-\epsilon$,
and $|\mathscr{T}_{\epsilon}^{k}| \leqslant 2^{k(\oH_1(X)+\epsilon)}$.
% The Conditionally Typical Set
A similar argument called joint AEP~\cite{Cover06B}
also holds for the i.i.d.\ sequences~$(X^k,Y^k)$ of length~$k$ drawn according to $\prod_{i=1}^k p_{X\,Y}(x_i,y_i)$.
Another information\hyph theoretic result is related to those sequences~$x^k$ that are jointly typical with a given typical sequence $y^k$.
Namely, the set of all these sequences~$x^k$ is referred to as the \emph{conditionally typical set} $\mathscr{T}_{\epsilon}^{X^k|y^k}$ and satisfies,
on the one hand, that $\oP\{\mathscr{T}_{\epsilon}^{X^k|y^k}\}> 1 - \epsilon$ for large $k$, and on the other,
that its cardinality is bounded by Shannon's conditional entropy, $|\mathscr{T}_{\epsilon}^{k}| \leqslant 2^{k(\oH_1(X|Y)+\epsilon)}$.
Further, it turns out that these conditionally typical sequences are equally likely, with probability $2^{-k\oH_1(X|Y)}$,
approximately in the exponent.
%Note, however, that these sequences are not necessarily the most likely ones.
While the most likely sequence may in fact \emph{not} belong to the typical set,
the set of typical sequences encompasses a sufficiently large number of sequences
that amount to a probability arbitrarily close to certainty.

Next, we proceed to interpret, under the perspective of our framework, the Shannon entropy as a measure of privacy.
To this end, consider the scenario in which a privacy attacker observes a typical~$Y^k$ and strives to estimate the unknown~$X^k$.
Conveniently, we assume $X^k=X'^k$ and $Y^k=Y'^k$, which models the LBS example described before,
provided that the attacker ignores any spatial\hyph temporal constraint.
In other words, we model a scenario without memory and hence suppose
that~$(X_i,Y_i)$ are i.i.d.\ drawn according to~$p_{X\,Y}$.
% Javi: Use the line below in my thesis.
%that~$(X_i,Y_i)$ are i.i.d.\ drawn according to~$p_{X\,Y}(x_i,y_i)$.
% 121002 Javi: In reply to Claudia's comment
We would like to stress that the consideration of this simplified model is just for the purpose of
providing a simple, clear example that illustrates the application of our framework.
Having said this, in the terms above we may regard $\mathscr{T}_{\epsilon}^{X^k|y^k}$ as a set of arbitrarily high confidence
with cardinality $2^{k\oH_1(X|Y)}$, approximately in the exponent.
%Having noted this, we contemplate an attacker model where, precisely, the attacker considers $\mathscr{T}_{\epsilon}^{X^k|y^k}$ as their confidence set.
%Since all sequences belonging to this set have a probability $2^{-k\oH_1(X|Y)}$ approximately in the exponent,
%a MAP strategy leads to
%$$\mathcal{P}(y^k)=1 - \max_{x^k} p(x^k|y^k) \approx 1 - 2^{-k \oH_1 (X|Y)},$$
%for a given a typical~$y^k$.
%To cut a long story short, our conditional privacy may be expressed in terms of the Shannon conditional entropy.

The upshot is that the Shannon (conditional) entropy of an unknown r.v.\ (given an observed r.v.) is an approximate measure of
the size of a high\hyph confidence set, measure suitable for attacker models based on the estimation of sequences,
rather than individual samples.
Moreover, within this confidence set, sequences are equally likely, approximately in the exponent,
concordantly with the interpretation of confidence\hyph set cardinality as a measure of privacy
made in Sec.~\ref{sec:Theory:Hamming:Single} on single occurrences.
Even though for simplicity our argument focused on memoryless sequences,
the Shannon\hyph McMillan\hyph Breiman theorem is a generalization of the AEP to stationary ergodic sequences,
in terms of entropy rates~\cite{Algoet88AP}.

Finally, we mentioned that the most likely sequence may in fact be atypical, and
thus Shannon entropy is not directly applicable to MAP estimation over the entire set of sequences.
Nevertheless, because the most likely memoryless sequence is simply a repetition of the most likely symbol,
MAP estimation on sequences is a trivial extension of the argument on min\hyph entropy presented in Sec.~\ref{sec:Theory:Hamming:Single}.
\subsection{Non\hyph Hamming Distortion}\label{sec:Theory:noHamming}
\noindent
This section investigates the complementary case described in Sec.~\ref{sec:Theory} in which the attacker's distortion function is not the Hamming distance.
Particularly, in this section we turn our attention to the scenario of SDC, and contemplate two possible alternatives regarding the system's knowledge on
the function $d_\textnormal{A}$---first, when this function is known to the data publisher, and secondly, when it is unknown.
Under the former assumption,
the system would definitely use BDT to find the decision rule~$p_{Y'|X'}$ which maximizes either the worst\hyph case privacy~\eqref{eq:worst_privacy}
or the average privacy~\eqref{eq:average_privacy},
and satisfies a constraint on average distortion.
The latter assumption, however, describes a more general and realistic scenario.
The remainder of this subsection precisely interprets several privacy criteria under this assumption.
The only piece of information which is though known to the publisher is~$d_\textnormal{max}=\max_{x,\hat{x}} d_\textnormal{A} (x,\hat{x})$,
that is, the \emph{maximum value} attained by said function.

\begin{figure*}
\centering
\includegraphics[scale=0.65]{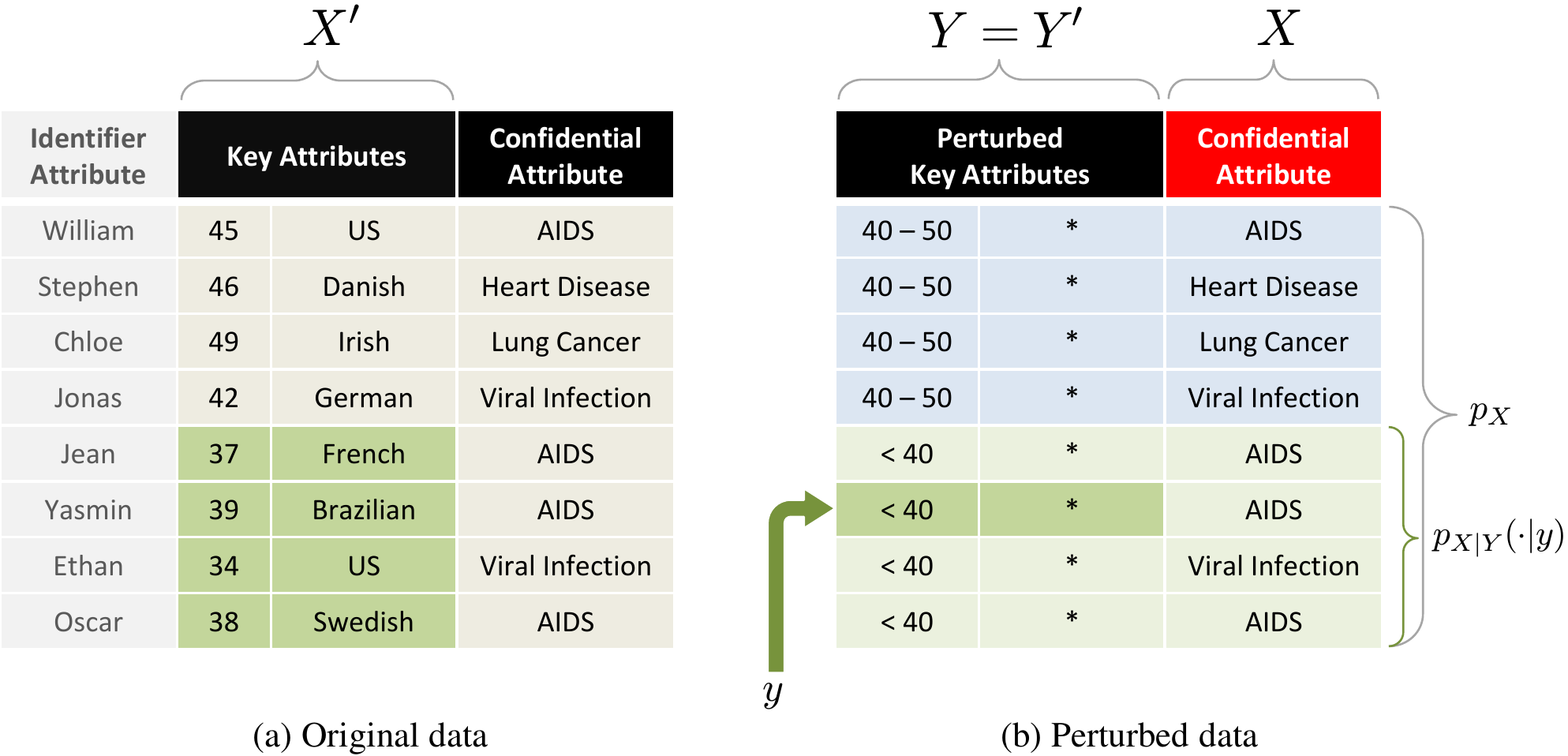}
\caption{At first, an attacker believes that the probability that a user appearing in (b) suffer from AIDS is 1/2.
However, after observing that the user's record is one of the last four records, this probability becomes 3/4.}
\label{fig:t-closeness}
\end{figure*}

Bearing in mind the above consideration, in our new scenario a privacy attacker endeavors to guess the confidential attribute value of a particular respondent in the released table.
Initially, the attacker has a prior belief given by~$p_X$, that is, the distribution of that confidential attribute value in the whole table.
Later, the attacker observes that the user belongs to a group of records sharing a tuple of perturbed key attribute values~$y$,
which is supposed to coincide with the system's decision~$y'$.
Based on this observation, the attacker updates their prior belief and obtains the posterior distribution~$p_{X|Y}(\cdot|y)$.
This situation is illustrated in Fig.~\ref{fig:t-closeness}.
A fundamental question that arises in this context is how much privacy the released table leaks as a result of that observation.
In the remainder of this section, we elaborate on this question and provide an upper bound on the reduction in privacy incurred by the disclosure of that information.

\SpaceBeforeSection

\subsubsection{Total Variation and $t$\hyph Closeness}

\noindent
For notational simplicity, we occasionally rename the posterior and the prior distributions $p_{X|Y}(\cdot|y)$ and $p_X$
simply with the symbols $p$ and~$q$, respectively, but bear in mind that $p$ is a PMF of $x$ parametrized by~$y$.
In addition, we shall assume that the attacker adopts a MAP strategy.
More precisely, $\hat{x}_p$ and $\hat{x}_q$ will denote the attacker's estimate when using the distributions $p$ and~$q$.
Under these assumptions, the \emph{reduction} (prior minus posterior) in conditional privacy can be expressed as
\begin{align*}
\Delta\mathcal{P}(y)  &= \oE_p d_\textnormal{A} (X,\hat{x}_q) - \oE_p d_\textnormal{A} (X,\hat{x}_p)\\
                      &= \oE_p d_\textnormal{A} (X,\hat{x}_q) - \oE_q d_\textnormal{A} (X,\hat{x}_q) + \oE_q d_\textnormal{A} (X,\hat{x}_q)\\
                      &- \oE_q d_\textnormal{A} (X,\hat{x}_p) + \oE_q d_\textnormal{A} (X,\hat{x}_p) - \oE_p d_\textnormal{A} (X,\hat{x}_p),
\end{align*}
where~$\oE_p$ and~$\oE_q$ denotes that the expectation is taken over the posterior and the prior distributions, respectively,
as PMFs of~$x$.

In this expression, the first two terms can be upper bounded by
$d_\textnormal{max} \sum_x \left|p_x -q_x\right|$, since $\sum_x (p_x - q_x)\leqslant \sum_x \left|p_x - q_x\right|$.
Clearly, this same bound applies to the last two terms.
On the other hand, the remaining terms $\oE_q d_\textnormal{A} (X,\hat{x}_q) -\oE_q d_\textnormal{A} (X,\hat{x}_p)$ are upper bounded by~0,
since the error incurred by~$\hat{x}_q$ is smaller than or equal to that of~$\hat{x}_p$.
In the end, we obtain that
$$\Delta\mathcal{P}(y) \leqslant 2\,d_\textnormal{max} \sum_x \left|p_x -q_x\right|.$$

At this point, we shall briefly review the concept of \emph{total variation}.
For this purpose, consider~$P$ and~$Q$ to be two PMFs over~$\mathcal{X}$.
In probability theory, the total variation distance between~$P$ and~$Q$ is
$$\textnormal{TV}(P\,\|\,Q) = \tfrac{1}{2} \sum_{x\in \mathcal{X}} \left|P(x) - Q(x)\right|.$$
Furthermore, recall that, in information theory, \emph{Pinsker's inequality} relates the total variation distance with the KL divergence.
Particularly, $\textnormal{TV}(P\,\|\,Q)\leqslant \tfrac{\sqrt{2}}{2} \sqrt{ \,\oD(P\,\|\,Q)}$.
Having stated this result, now the total variation distance permits writing the upper bound on~$\Delta\mathcal{P}(y)$ in terms of the KL divergence:
$$\Delta\mathcal{P}(y) \leqslant 4\,d_\textnormal{max} \textnormal{TV}(p\,\|\,q) \leqslant 2\sqrt{2}\, d_\textnormal{max}\, \sqrt{\oD(p\,\|\,q)},$$
where the last inequality follows from Pinsker's inequality.
Returning to the notation of prior and posterior distributions,
\begin{multline}\label{eq:upperbounds}
\Delta\mathcal{P}(y) \leqslant 4\,d_\textnormal{max} \textnormal{TV}(p_{X|Y}(\cdot|y)\,\|\,p_X)\\
\leqslant 2\sqrt{2} \, d_\textnormal{max}\, \sqrt{\oD(p_{X|Y}(\cdot|y)\,\|\,p_X)}.
\end{multline}

This upper bound allows to establish a connection between our privacy criterion and $t$\hyph closeness~\cite{Li07ICDE}.
The latter criterion boils down to defining a maximum discrepancy between the posterior and prior distributions,
$$t=\max_y \oD (p_{X|Y}(\cdot|y)\,\|\,p_X).$$
Under this definition and on account of~(\ref{eq:upperbounds}),
$$\Delta\mathcal{P}(y)\leqslant 2\sqrt{2} \, d_\textnormal{max}\,    \sqrt{t}.$$
Therefore, $t$\hyph closeness is essentially equivalent to bounding the decrease in conditional privacy.

On a different note, we would like to make a comment on an issue of a purely technical nature.
Clearly, in light of inequality~(\ref{eq:upperbounds}), the minimization of either the total variation distance or the KL divergence leads to the
minimization of an upper bound on~$\Delta \mathcal{P}(y)$.
However, the fact that the KL divergence imposes a worse upper bound suggests us considering it when the resulting mathematical model be
more tractable than the one built upon the total variation distance.

\SpaceBeforeSection

\subsubsection{Mutual Information and Rate\hyph Distortion Theory}

\noindent
The privacy criterion proposed in~\cite{Rebollo10KDE}, called \emph{(average) privacy risk}~$\cR$,
is the average\hyph case version of $t$\hyph closeness.
Formally, $\cR$ is a conditional KL divergence, the average discrepancy between the posterior and the prior distributions,
which turns out to coincide with the mutual information between the confidential data $X$ and the observation~$Y$:
\begin{multline*}
\cR=\oE_Y \oD (p_{X|Y}(\cdot|Y)\,\|\,p_X)\\
=\oE_Y \oE_{X|Y} \left[\left. \log \frac{p_{X|Y}(X|Y)}{p_X(X)}\right|Y \right]\\
=\oE \log \frac{p_{X|Y}(X|Y)}{p_X(X)} = \oI(X;Y).
\end{multline*}
Directly from their definition, $\cR\leqslant t$, meaning that $t$\hyph closeness is a stricter measure of privacy risk.
Because the KL divergence is itself an average,
$\cR$ is clearly an average\hyph case privacy criterion, but $t$ closeness is technically a maximum of an expectation,
a hybrid between average case and worst case.
The next subsection will comment on a third, purely worst\hyph case criterion.
When choosing a privacy criterion, it is important to keep in mind that
optimizing a privacy mechanism for the best worst\hyph case scenario will in general yield a worse average case, and viceversa.

Further, we conveniently rewrite inequality~(\ref{eq:upperbounds}) as
$$\tfrac{1}{8\,d^2_\textnormal{max}}\Delta \mathcal{P}(y)^2 \leqslant \oD (p_{X|Y}(\cdot|y)\,\|\,p_X).$$
By averaging over all possible observation~$y$, the right\hyph hand side of this inequality becomes
the privacy risk~$\cR$, which we showed to be equal to the mutual information.
This leads to a bound on the privacy reduction in terms of mutual information,
$$\tfrac{1}{8\,d^2_\textnormal{max}} \oE \left[\Delta \mathcal{P}(Y)^2 \right]\leqslant \oI(X;Y).$$

Based on this observation, it is clear that the minimization of the mutual information contributes to the minimization of
an upper bound on~$\Delta \mathcal{P}(y)$.
With this in mind, we now consider the more general scenario in which $Y'$ and $Y$ need not necessarily coincide, and contemplate the case of a data publisher.
Concretely, from the perspective of a publisher, we would choose a randomized perturbation rule~$p_{Y'|X'}$ with the aim of minimizing the mutual information between $X$ and~$Y$, and consequently protecting user privacy.
Evidently, the publisher would also need to guarantee the utility of the data to a certain extent, and thus impose a constraint on the average distortion.
In conclusion, the data publisher would strive to solve the optimization problem
\begin{equation}\label{eq:ratedistortion}
\min_{\substack{p_{Y'|X'}\\\oE d_\textnormal{U}(X',Y')\leqslant\cD}} \oI(X;Y),
\end{equation}
which surprisingly bears a strong resemblance with the rate\hyph distortion problem in the field of information theory.

More specifically, the above optimization problem is a generalization of a well\hyph known, extensively studied information\hyph theoretic problem
with more than half a century of maturity.
Namely, the problem of lossy compression of source data with a distortion criterion, first proposed by Shannon in 1959~\cite{Shannon59IRE}.

The importance of this lies in the fact that some of the information\hyph theoretic results and methods for the rate\hyph distortion problem can
be extended to the problem~(\ref{eq:ratedistortion}).
For example, in the special case when~$X=X'$ and $Y=Y'$, our more general problem boils down to Shannon's rate\hyph distortion and, interestingly,
can be computed with the Blahut\hyph Arimoto algorithm~\cite{Cover06B}.

Bear in mind that the very same metric, or conceptually equivalent variations thereof, may in fact be interpreted under
different perspectives.
Recall, for instance, that mutual information is the difference between an unconditional entropy and a conditional entropy,
effectively the posterior uncertainty modeled simply by the Shannon entropy, normalized with respect to its prior correspondence.
Under this perspective, mutual information might also be connected to the branch of the tree in Fig.~\ref{fig:sketch} leading to Shannon's entropy.

\SpaceBeforeSection

\subsubsection{$\delta$\hyph Disclosure and Differential Privacy}

Finally, we quickly remark on the connection of $\delta$\hyph disclosure
and $\epsilon$\hyph differential privacy with our theoretical framework.
$\delta$\hyph \emph{disclosure}~\cite{Brickell08KDD} is an even stricter privacy criterion than $t$\hyph closeness, and hence much stricter than that average privacy risk $\cR$ or mutual information, discussed in the previous subsection.
The definition of $\delta$\hyph disclosure may be rewritten in terms of our notation as
$$\delta=\max_{x,y} \left| \log \frac{p_{X|Y}(x|y)}{p_X(x)} \right|,$$
and understood as a worst\hyph case privacy criterion.
In fact,
$$\cR\leqslant t\leqslant \delta.$$

We mentioned in the background section that~\cite{Dwork06A}
analyzes the case of the randomized perturbation $Y$ of a true answer $X$ to a query in a private information retrieval system,
before returning it to the user.
Consider two databases $d$ and $d'$ that differ only by one record, but are subject to a common perturbation rule $p_{Y|X}$,
and let $p_Y$ and $p'_Y$ be the two probability distributions of perturbed answers induced.
After a slight manipulation of the definition given in the work cited, but faithfully to its spirit,
we may say that a randomized perturbation rule provides
$\epsilon$\hyph differential privacy when
$$\epsilon=\max_{y,d,d'} \log \frac{p_Y(y)}{p'_Y(y)}.$$
Even though it is clear that this formulation does not quite match
the problem in terms of prior and posterior distributions described thus far,
this manipulation enables us to still establish a loose relation with $\delta$\hyph disclosure,
in the sense that the latter privacy criterion is a slightly stricter measure of discrepancy between PMFs,
also based on a maximum (absolute) log ratio. We note, however, that although there is a formal similarity between the metrics, there are substantial differences between them in terms of their assumptions, objectives, models, and privacy guarantees.

%%%%%%%%%%%%%%%%%%%%%%%%%%%%%%%%%%%%%%%%%%%%%%%%%%%%%%%%%%%%%%%%%%%%%%%%%%%%%%%%%%%%%%%%%%%%%%%%%%%%%%%%%%%%%%%%%%%%%%%%%%%%%%%%%%
%6 Numerical Examples
%%%%%%%%%%%%%%%%%%%%%%%%%%%%%%%%%%%%%%%%%%%%%%%%%%%%%%%%%%%%%%%%%%%%%%%%%%%%%%%%%%%%%%%%%%%%%%%%%%%%%%%%%%%%%%%%%%%%%%%%%%%%%%%%%%
\section{Numerical Example}\label{sec:Numerical}
\noindent
%Intro
This section provides two simple albeit insightful examples that illustrate the measurement of privacy as an attacker's estimation error.
Specifically, we quantify the level of privacy provided,
first, by a privacy\hyph enhancing mechanism that perturbs location information in the scenario of LBS, and secondly, by
an anonymous\hyph communication protocol largely based on Crowds~\cite{Reiter98ISS}.

\subsection{Data Perturbation in Location\hyph Based Services}\label{sec:Numerical:LBS}
\noindent
Our first example contemplates a user who wishes to access an LBS provider.
For instance, this could be the case of a user who wants to find the closest Italian restaurant to their current location.
For this purpose, the user would inevitably have to submit their GPS coordinates to the (untrusted) provider.
To avoid revealing their exact location, however,
the user itself could perturb their location information by adding, for example, Gaussian noise.
Alternatively, we could consider a user delegating this task to a (trusted) intermediary entity, as described in Sec.~\ref{sec:Theory:Hamming:Multiple}.
%121002 Claudia: Remove this.
%thus yielding what we called soft privacy in Sec.~\ref{sec:Formulation:Notation}.
In any case, data perturbation would enhance user privacy in terms of location,
although clearly at the cost of data utility.
Simply put, perturbative privacy methods present the inherent trade\hyph off between data utility and privacy.

\begin{figure}[htb]
\centering
\includegraphics[scale=0.55]{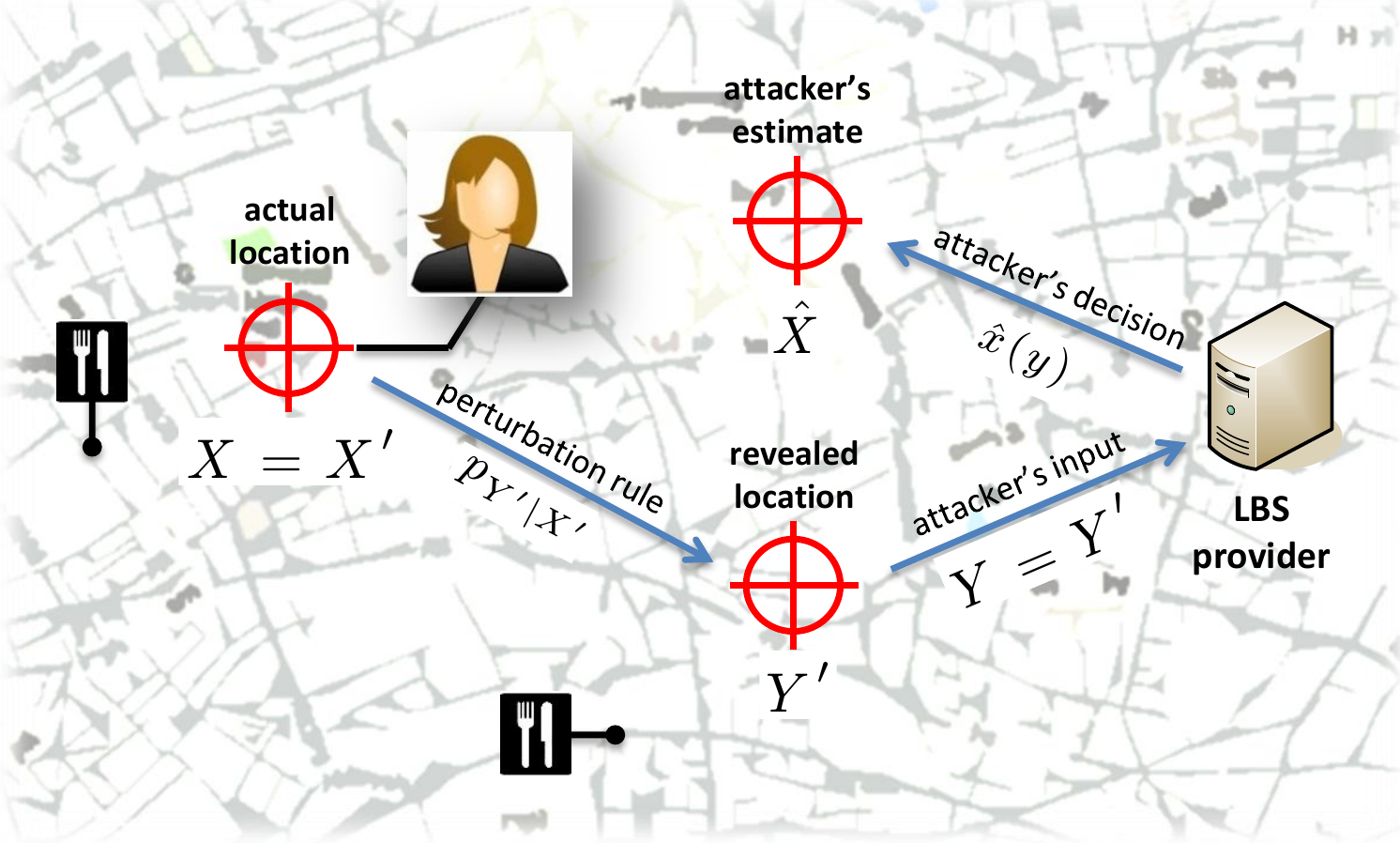}
\caption{A user looking for a nearby Italian restaurant accesses an LBS provider.
The user decides to perturb their actual location before querying the provider.
In doing so, the user hinders the provider itself and any attacker
capable of capturing their query, in their efforts to compromise user privacy in terms of location.
% 121002 Claudia: Remove the concepts of hard privacy and soft privacy.
%In this example, we contemplate a hard\hyph privacy approach---the user is solely responsible for protecting their private data.
In this example, we contemplate that the user is solely responsible for protecting their private data.
In terms of our notation, this allows us to regard the user as the system.
Notice that the user's actual location is, on the one hand, the attacker's unknown, and on the other, the information that the user (system) takes as input
to generate the location that will be finally revealed. Thus we conclude that $X=X'$.
Then, according to some randomized perturbation rule $p_{Y'|X'}$, the user discloses, for each location data $x'$, a perturbed version~$y'$.
This perturbed location is submitted to the provider, which only has access to this information, i.e., $Y=Y'$.
Lastly, based on this revealed information, the attacker uses a Bayes estimator $\hat{x}(y)$ to ascertain the user's actual location~$X$.}
\label{fig:LBS}
\end{figure}
%121002 Claudia: Remove the concepts of hard privacy and soft privacy.
%Under the former strategy, i.e., that providing hard privacy, and in accordance with the notation defined in Sec.~\ref{sec:Formulation:Notation},
Under the former strategy, and in accordance with the notation defined in Sec.~\ref{sec:Formulation:Notation},
the user becomes the system---it is the user who is responsible for protecting their location data.
Playing the role of the system, the user decides then to perturb their location data $X$ on an individual basis for each query.
In other words, we do not contemplate the case of sequences of data $X^k$, as Sec.~\ref{sec:Theory:Hamming:Multiple} does.

A key element of our framework is the attacker's distortion function. In our example we assume the squared error between the actual location $x$ and the attacker's estimate $\hat{x}$, that is, $d_\textnormal{A}(x,\hat{x}) =  \| x - \hat{x} \|^2$.
Unlike Hamming distance, note that the squared error does quantify how much the estimate differs from the unknown.
As for the other variables of our model, we contemplate that the attacker's input $Y$ is directly the location data perturbed by the user, $Y'$,
as illustrated in Fig.~\ref{fig:LBS}.
Put differently, the attacker, assumed to be the service provider, has no more information than that disclosed by the user.
Under all these assumptions, the average privacy~\eqref{eq:average_privacy} is
$$\mathcal{P}_\textnormal{avg} = \oE[\|X - \hat{X}\|^2],$$
that is, the mean squared error (MSE).

As a final remark, we would like to connect our privacy criterion with a metric specifically conceived for the LBS scenario at hand~\cite{Shokri11SP}.
In this cited work, the authors propose a framework that contemplates different aspects of the adversarial model,
captured by means of what they call \emph{certainty}, \emph{accuracy} and \emph{correctness}.
The information to be protected by a trusted intermediary system are traces modeling the locations visited by users over a period of time.
The system accomplishes this task by hiding certain locations, reducing the accuracy of such locations or adding noise.
As a result, the attacker observes a perturbed version of the traces and, together with certain mobility profiles of these users,
attempts to deduce some information of interest~$X$ about the actual traces.
In terms of our notation, the observed trajectories and the mobility patterns constitute the attacker's observation~$Y$.

More accurately, given a particular observation~$y$, the attacker strives to calculate the posterior distribution~$p_{X|Y}$.
However, since the adversary may have a limited number of resources, they may have to content themselves with an estimate~$\hat{p}_{X|Y}$.
The authors then use Shannon's entropy to measure the \emph{uncertainty} of~$X$,
and define \emph{accuracy} as the discrepancy between $p_{X|Y}$ and $\hat{p}_{X|Y}$.
Finally, they refer to location privacy as \emph{correctness} and measure it as
$$\textnormal{E}_{\hat{p}_{X|Y}} [d_{\textnormal{S}}(X,x_t)|y],$$
where $x_t$ is the true outcome of $X$, $d_{\textnormal{S}}$ a distance function specified by the system,
and the expectation is taken over the estimate of the posterior distribution.

The most notable difference between~\cite{Shokri11SP} and our own work is that the authors limit the scope of their metric to the specific scenario of location\hyph based services; whereas here we attempt to provide a general overview.
Besides, their proposal is a measure of privacy in an average\hyph case sense.
%Another difference is that they argue against the use of entropy and $k$\hyph anonymity for the purposes of their field of application.
%In this regard, we not only justify but also relate entropy and $k$\hyph anonymity within a generalized perspective on attacker's estimation errors, rather than excluding them.
Another important distinction between the cited work and ours is that the former arrives to the conclusion that entropy and \emph{k}\hyph anonymity are not appropriate metrics for quantifying privacy in the context of LBS.
Our work, however, does \emph{not} argue against the use of entropy, \emph{k}\hyph anonymity and any of the other privacy metrics examined in Sec.~\ref{sec:Theory}. In fact, we regard these metrics as particular cases of the attacker's estimation error under certain assumptions on the adversarial model, the attacker's strategy and a number of different considerations explored in that section.
Lastly, their implementation of estimation strategies using the forward\hyph backward~\cite{Reid79AC} and the Metropolis\hyph Hastings~\cite{Hastings70Bio} algorithms are undoubtedly of great interest, but the focus of the present work is on metrics.

\subsection{Crowds\hyph like Protocol for Anonymous Communications}\label{sec:Numerical:Crowds}
\noindent
In Sec.~\ref{sec:Background:ACS_LBS} we mentioned Chaum's mixes as a building block to implement anonymous communications networks.
A different approach to communication anonymity is based on collaborative, peer-to-peer architectures.
An example of collaborative approach is Crowds~\cite{Reiter98ISS}, in which users form a ``crowd" to provide anonymity for each other.

In Crowds, a user who wants to browse a Web site forwards the request to another member of his crowd chosen uniformly at random.
This crowd member decides with probability $p$ to send the request to the Web site, and with probability $1-p$ to send it to another randomly chosen crowd member,
who in turn repeats the process.
For the purposes of illustration, we consider a variation of the Crowds protocol.
The main difference with respect to the original Crowds is that we do not introduce a mandatory initial forwarding step.
We note that this variation provides worse anonymity than the original protocol, while also reducing the cost (in terms of delay and bandwidth) with respect to Crowds.
Further, we assume that the users participating in the protocol are honest; i.e., we only consider the Web site receiving the request as possible adversary.

\begin{figure}[htb]
\centering
\includegraphics[scale=0.55]{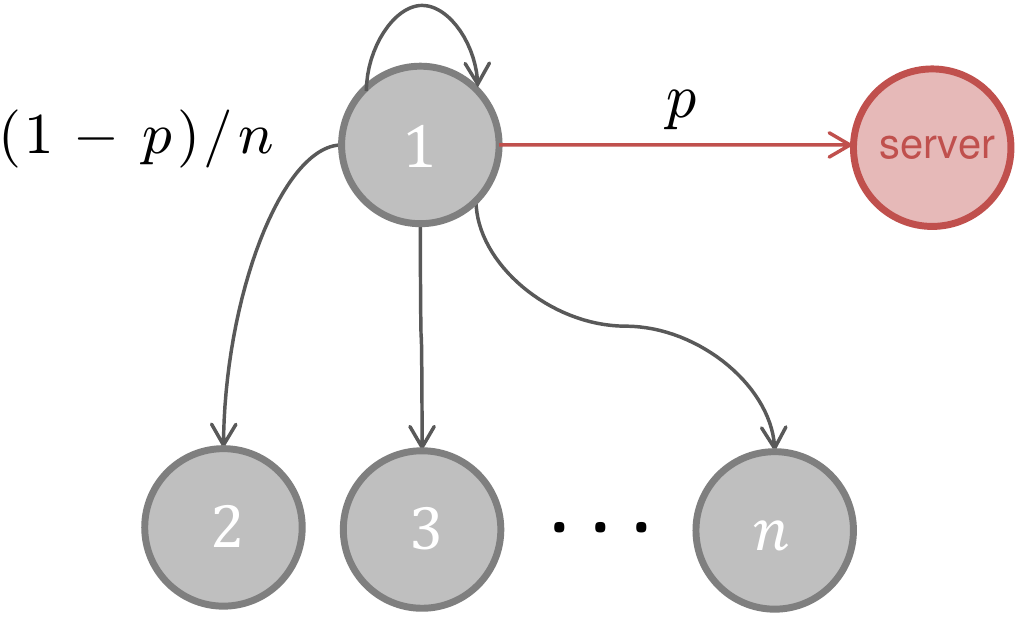}
\caption{Anonymous\hyph communication protocol inspired by Crowds.
In our second numerical example, we contemplate a scenario where users send messages to a common, untrusted server,
who aims at compromising sender anonymity.
In response to this privacy threat,
users decide to adhere to a modification of the Crowds protocol,
whose operation is as follows:
each user flips a biased coin and depending
on the outcome chooses to submit the message to the
server or else to another user, who is asked to perform
the same process.
The probability that a user forward the message to the server is denoted by $p$,
whereas the probability of sending it to any other peer, including themselves, is $(1-p)/n$.}
\label{fig:Crowds}
\end{figure}

More formally, consider $n$ users indexed by $i=1,\ldots, n$, wishing to communicate with an untrusted server.
In order to attain a certain degree of anonymity, each user submits the message directly to
said server with probability $p\in(0,1)$, and forwards it to any of the other users, including themselves, with probability $1-p$.
In the case of forwarding, the recipient performs exactly the same probabilistic decision until the message arrives at the server.
Fig.~\ref{fig:Crowds} shows the operation of this protocol.

In our protocol, we assume that the server attempts to guess the identity of the author of a given message, represented by the r.v.\ $X$,
knowing only the user who last forwarded it, represented by the r.v.\ $Y$, consistently with the notation defined in Sec.~\ref{sec:Formulation:Notation}.
The other variables of our framework are as follows.
Since the set of users involved in the protocol collaborate to frustrate the efforts of the server, they are in fact the system.
The information that then serves as input to this system is simply the identity of the user who initiates the forwarding protocol, $X$.
That is, the attacker's uncertainty and the system's input coincide, $X'=X$.
Then again, the assumption that the server just knows the last sender in the forwarding chain leads to~$Y=Y'$.

\begin{figure}[htb]
\centering
\includegraphics[scale=0.95]{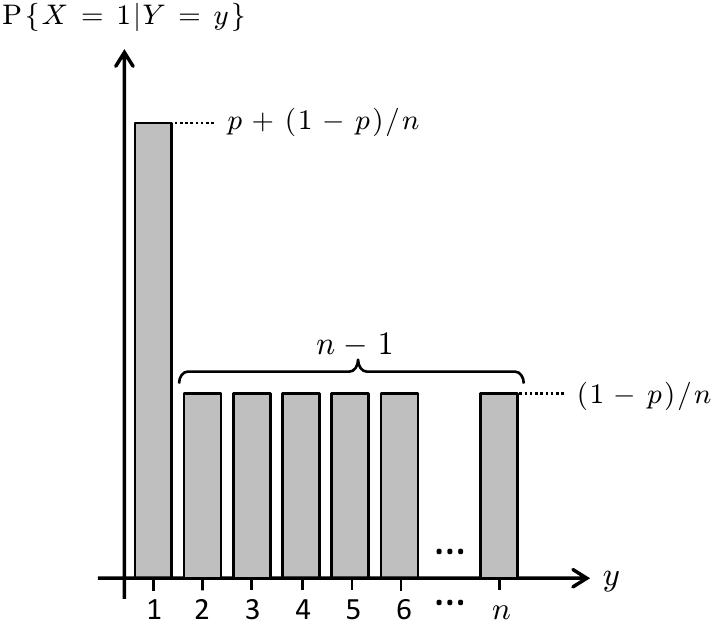}
\caption{Probability that the original sender of a given message be the user 1,
conditioned to the observation that the last sender in the forwarding path is user $y$.
From this figure, we observe the PMF attains its maximum value when this last sender is precisely the user 1.}
\label{fig:PMFCrowds}
\end{figure}

Under this model, and under the assumption of a uniform message\hyph generation rate, that is, $p_X(x)=1/n$ for all~$x$,
it can be proven that the conditional PMF of $X$ given $Y=y$ is
\begin{equation}\label{eq:PMF}
p_{X|Y}(x|y)=\left\{ \begin{array}{l@{,\quad}l}
                                            p + (1-p)/n \quad \quad & x=y\\
                                            (1-p)/n \quad \quad & x \neq y
                \end{array}\right..
\end{equation}
Fig.~\ref{fig:PMFCrowds} shows this conditional probability in the particular case when $x=1$, i.e., the probability that the originator of a message be user 1, conditioned to the observation that the last sender is user~$y$.
Note that, because of the symmetry of our model, it would be straightforward to derive a PMF analogous to the one plotted in this figure, but for other originators of the message, namely $x=2,\ldots,n$.

That said, assume that the attacker chooses Hamming distance as distortion function.
Under this assumption, the conditional privacy~\eqref{eq:conditional_privacy} yields
$$\mathcal{P}(y) = \oP\{X \neq\hat{x}(y)|y\},$$
that is, the MAP error conditioned on the observation~$y$.
Because Hamming distance implies, by virtue of~\eqref{eq:BayesMAP}, that Bayes estimation is equivalent to MAP estimation,
it follows that the attacker's (best) decision rule is $\hat{x}(y)=y$.
Leveraging on this observation, we obtain that the privacy level provided by this variant of Crowds is
$$\mathcal{P}(y) = \varepsilon_{_\textnormal{MAP}} = 1 - \oP\{X=y|y\} = (1-p)(1-1/n),$$
from which it follows an entirely expected result---the lower the probability $p$ of forwarding a message directly to the server, the higher the privacy provided by the protocol, but the higher the delay in the delivery of said message.

In the following, we consider the measurement of the privacy protection offered by this protocol,
in terms of the three R\'enyi's entropies introduced in Sec.~\ref{sec:Theory:Hamming}, namely the min\hyph entropy $\oH_\infty(X|y)$, the Shannon entropy $\oH_1(X|y)$ and the Hartley entropy $\oH_0(X|y)$ of the r.v.~$X$, modeling the actual sender of a given message (the privacy attacker's target),
given the observation of the user who last forwarded it,~$y$.
Specifically, we connect the interpretations described in Sec.~\ref{sec:Theory:Hamming} to the example at hand.

But first we would like to recall from Sec.~\ref{sec:Theory:Hamming:Single}
that $\oH_\infty(X|y)$, $\oH_1(X|y)$ and $\oH_0(X|y)$ may be considered, from the point of view of the user, as a worst\hyph case, average\hyph case and best\hyph case measurements of privacy, respectively, in the sense that
$$\oH_\infty(X|y) \leqslant \oH_1(X|y) \leqslant \oH_0(X|y),$$
owing to~\eqref{eqn:Theory:EntropyInequalities}, with equality if and only if the conditional PMF of $X$ given $Y=y$ is uniform.
Revisiting the interpretations given in that section, recall that the min\hyph entropy $\oH_\infty(X|y)$ is directly connected with the maximum probability, in our case $\max_{x_i} p_{X|Y} (x_i|y) =p + (1-p)/n$, on account of~\eqref{eq:PMF}.
More concretely, and in the context of our example, min\hyph entropy reflects the model in which a privacy attacker makes a single guess of the originator of a message, specifically the most likely one, which corresponds to $x=y$.

At the other extreme, the Hartley entropy $\oH_0(X|y)$ is a possibilistic rather than probabilistic measure, as it corresponds to
the assumption that a privacy attacker would not content themselves with discarding all but the most likely sender,
but consider instead all possible users.
More accurately, measuring privacy as a Hartley's entropy essentially boils down to the cardinality of the set of all possible originators of a message,
namely $\oH_0 (X|y) = \log n.$

On a middle ground lies Shannon's entropy, which was interpreted in Sec.~\ref{sec:Theory:Hamming:Multiple} by means of the AEP,
specifically in terms of the effective cardinality of the set of typical sequences of i.i.d.\ samples of an r.v.
Put in the context of our Crowds\hyph like protocol, however, Shannon's entropy may be deemed as an average\hyph case metric that considers the entire PMF of~$X$ given $Y=y$, and not merely its maximum value or its support set.

%%%%%%%%%%%%%%%%%%%%%%%%%%%%%%%%%%%%%%%%%%%%%%%%%%%%%%%%%%%%%%%%%%%%%%%%%%%%%%%%%%%%%%%%%%%%%%%%%%%%%%%%%%%%%%%%%%%%%%%%%%%%%%%%%%
%7 Guide for System Designers
%%%%%%%%%%%%%%%%%%%%%%%%%%%%%%%%%%%%%%%%%%%%%%%%%%%%%%%%%%%%%%%%%%%%%%%%%%%%%%%%%%%%%%%%%%%%%%%%%%%%%%%%%%%%%%%%%%%%%%%%%%%%%%%%%%
\section{Guide for Designers of SDC and ACSs}\label{sec:Guide}
\noindent
The purpose of this section is to show the applicability of our framework to those designers of SDC and ACSs who,
wishing to quantify the level of protection offered by their systems, do not want to delve into the mathematical details set forth in Sec.~\ref{sec:Theory}.
%serve as a guide for those system designers who wish to quantify the level of protection offered by
%their privacy\hyph enhancing technologies, without having to delve into the mathematical details set forth in Sec.~\ref{sec:Theory}.
In order to assist such designers in the selection of the privacy metric most appropriate for their requirements,
this section revises the application scenarios of SDC and anonymous communications,
and classifies some of the metrics used in these fields in terms of worst case, average case and best case, from the perspective of the user.

Before proceeding with the cases of SDC and ACSs, we would like to elaborate on the distinction between Hamming and non\hyph Hamming distortion functions, between whether these functions are known or unknown to the system, and finally between single and multiple-occurrence data. The reason is that the understanding of these concepts is fundamental for a system designer who, following the arguments sketched in Fig.~\ref{fig:sketch},
wants to choose the suitable metrics for their field of application. With this purpose, next we illustrate these concepts by means of a couple of simple albeit insightful examples.

The first consideration a system designer should take into account when applying our framework refers to the geometry of the attacker's distortion function~$d_{\textnormal{A}}$, namely whether it is a \emph{Hamming} or a\emph{ non-Hamming} function. To illustrate this key point, consider a set of users in a social network. A Hamming function taking as inputs the users $u_1$ and $u_2$ would model an attacker who contemplates \emph{only} their identities when comparing them, and ignores any other information such as the relationship between them within the social network, their profile similarity or their common interests. On the contrary, a more sophisticated adversary could represent said network by a graph, modeling users and relationships among them as nodes and edges, respectively. Leveraging on this graph, the attacker could use a non-Hamming function to compute the number of hops separating these two users and, accordingly, lead to the conclusion that they are, for example, close friends since $d_{\textnormal{A}}(u_1,u_2) = 1$.

The second consideration builds on the assumption of a non-Hamming attacker's distortion function. Under this premise, we contemplate two possible cases---when the function is \emph{known} to the system and when \emph{not}. The former case is illustrated, for instance, in the context of location-based services---in this application scenario, an adversary will probably use the Euclidean distance to measure how their estimated location differs from the user's actual location. The latter case, i.e., when the measure of distortion used by the attacker is unknown to the system, would undoubtedly model a more general and realistic scenario. As an example of this case, consider a system perturbing the queries that a user wants to submit to a database, and an attacker wishing to ascertain the actual queries of this user. Suppose that these queries are one-word queries and that the perturbation mechanism replaces them with synonyms or semantically-similar words. Under these assumptions, our attacker could opt for a non-Hamming distortion function and measure the distance between the actual query and the estimate as the number of edges in a given ontology graph. Although the system could be aware of this fact, the specific ontology used by the attacker could not be available to the system, and consequently the distortion function would remain unknown.

Our last consideration is related to the nature of the variables of our framework, summarized in Table~\ref{tab:notation:general}. Specifically, we contemplate two possible cases---\emph{single} and \emph{multiple-occurrence} data. The former case considers such variables to be tuples of a small number of components, and the latter assumes that these variables are sequences of data. An LBS  attacker who observes the disclosed, possibly perturbed location of a user and makes a single guess about their actual location is an example of single-occurrence data. To illustrate the case of multi-occurrence data, consider a set of users exchanging messages through a mix system. Recall that such systems delay and reorder messages with the aim of concealing who is communicating with whom. Among the multiple attacks these systems are vulnerable to, the statistical disclosure attack~\cite{Danezis03SEC} is a good example for our purposes of illustration,
since it assumes an adversary who observes a large number or \emph{sequence} of messages coming out of the mix, with the aim of tracing back their originators.
%wishes to trace back the of the messages coming out of the mix by observing a large number or \emph{sequence} of these outgoing messages.

Having examined these key aspects of our framework, now we turn our attention, first, to the application scenario of SDC, and secondly, to the case of ACSs.
% SDC
In the former scenario, a data publisher aims at protecting the privacy of the individuals appearing in a microdata set.
Depending on the privacy requirements, the publisher may want to prevent an attacker from ascertaining the confidential attribute value of any respondent in the released table.
Under this requirement, $t$\hyph closeness and mutual information appear as acceptable measures of privacy,
since both criteria protect against confidential \emph{attribute disclosure}.
Recall that the assumptions on which they are based are a prior belief about the value of the confidential attribute in the table,
and a posterior belief of said value given by the observation that the user belongs to a particular group of this table.
Building on these premises, $t$\hyph closeness may be regarded as a \emph{worst\hyph case} measurement of privacy,
in the sense that it identifies the group of users whose distribution of the confidential attribute deviates the most from the distribution of this same attribute in the entire table.
In this regard, we would like to note that a worst\hyph case metric from the point of view of the user is a best\hyph case measure from the standpoint of the attacker,
and vice versa.

Although $t$\hyph closeness overcomes the similarity and skewness attacks mentioned in Sec.~\ref{sec:Background:SDC},
its main limitation is that no computational procedure to reach this criterion has been specified.
An alternative is the mutual information between the confidential attributes and the observation,
an average\hyph case version of $t$\hyph closeness that leads to a looser measure of privacy.
In any of these two metrics, it is assumed the more general case in which the attacker's distortion function is not the Hamming distance.
Specifically, this assumption models an adversary who does not content themselves with finding out whether the estimate and the unknown match,
but wishes to quantify how much they diverge.

Another distinct privacy requirement is that of \emph{identity disclosure},
whereby a publisher wishes to protect the released table against a linking attack.
In this attack, the adversary's aim is to uncover the identity of the individuals
in the released table by linking the records in this table to a public data set including identifier attributes.
Under this requirement and under the assumption that the attacker regards each respondent within a particular group as equally likely,
$k$\hyph anonymity may be deemed as a \emph{best\hyph case} measure of privacy, determined by Hartley's entropy.
We refer to this criterion as a best\hyph case metric precisely due to the naive assumption of a uniform distribution of the identifier attribute.
In other words, the underlying adversarial model does not contemplate that an attacker may have background knowledge that allows them to consider certain users as more likely than others.
In the end, we may also regard the $l$\hyph diversity criterion as a best\hyph case metric,
since it assumes a uniform distribution of the confidential attribute on a set of at least $l$ values.
Put another way, this rudimentary adversarial model does not contemplate, for example, the fact that certain values of the confidential attribute may be semantically similar.

% ACS
In the scenario of anonymous\hyph communication systems, there exists a wide variety of approaches.
%121002 Claudia: Remove the phrase below.
%many of them relying on the assumptions of soft privacy.
Among them, a popular anonymous\hyph communication protocol is Crowds.
Although in this section we limit the discussion of the privacy provided by such systems to a variant of this protocol,
we would like to stress that the conclusions drawn here may be extended to other anonymous systems.
Having said this, recall that in the original Crowds protocol, a system designer
makes available to users a collaborative protocol that helps them enhance the anonymity of the messages sent to a common, untrusted Web server.
The design parameters are the number of users participating in the protocol and the probability of forwarding a message directly to the server.

In our variant of this protocol, however, we contemplate an attacker who strives to guess the identity of the sender of a given message,
based on the knowledge of the last user in the forwarding path.
Under this adversarial model, we may regard min\hyph entropy, Shannon's entropy or Hartley's entropy as particular cases of our measure of privacy,
depending on the specific strategy of the attacker.
For example, under an adversary who uses maximum a posteriori estimation and, accordingly, opts for the last sender,
min\hyph entropy may be interpreted as a worst\hyph case privacy metric.
Alternatively, we may assume an attacker that considers the entire probability distribution of possible senders, and not only the most likely candidate.
In this case, Shannon's entropy may be deemed as an average\hyph case measure.
Finally, under a rudimentary attacker who takes into account just the number of potential originators of the message,
Hartley's entropy may be regarded as a best\hyph case measurement of privacy. %synonym of deem

% LBS
%In the end, we revise the interesting case of LBS through the numerical example presented in Sec.~\ref{sec:Numerical:LBS}.
%In this example, we contemplate an attacker who endeavors to estimate the actual location of a user,
%based on observed, perturbed location data.
%In addition, we assume that the attacker's distortion function is the squared error between the actual location and the estimate,
%and that this function is known to the system designer.
%Under these premises, the mean squared error, that is, the expected value (average) of the squared error loss, boils down to the attacker's estimation error, i.e., the privacy metric proposed in this work. Table~\ref{fig:Guide} summarizes this section.

%%%%%%%%%%%%%%%%%%%%%%%%%%%%%%%%%%%%%%%%%%%%%%%%%%%%%%%%%%%%%%%%%%%%%%%%%%%%%%%%%%%%%%%%%%%%%%%%%%%%%%%%%%%%%%%%%%%%%%%%%%%%%%%%%%
%7 Conclusions
%%%%%%%%%%%%%%%%%%%%%%%%%%%%%%%%%%%%%%%%%%%%%%%%%%%%%%%%%%%%%%%%%%%%%%%%%%%%%%%%%%%%%%%%%%%%%%%%%%%%%%%%%%%%%%%%%%%%%%%%%%%%%%%%%%
\section{Conclusion}\label{sec:Conclusions}
\noindent
A wide variety of privacy metrics have been proposed in the literature. Most of these metrics have been conceived for specific applications, adversarial models, and privacy threats, and thus are difficult to generalize. Even for specific applications, we often find that various privacy metrics are available. For example, to measure the anonymity provided by anonymous\hyph communication networks, several flavors of entropy (Shannon, Hartley, min\hyph entropy) can be found in the literature, while no guidelines exist that explain the relationship between the different proposals, and provide an understanding of how to interpret and put in context the results provided by each of them.

In the scenario of SDC, numerous approaches attempt to capture, to a greater or lesser degree, the private information leaked as a result of the dissemination of microdata sets. In this spirit, $k$\hyph anonymity is possibly the best\hyph known privacy measure, mainly due to its mathematical tractability. However, numerous extensions and enhancements were introduced later with the aim of overcoming its limitations. While all these metrics have provided further insight into our understanding of privacy, the research community would benefit from a framework embracing all those metrics and making it possible to compare them, and to evaluate any privacy\hyph protecting mechanism by the same yardstick.

In this work, we propose a privacy measure intended to tackle the above issues.
Our approach starts with the definition and modeling of the variables of a general framework.
Then, we proceed with a mathematical formulation of privacy, which essentially emerges from BDT.
Specifically, we define privacy as the estimation error incurred by an attacker.
We first propose what we refer to as conditional privacy, meaning that our measure is conditioned on an attacker's particular observation.
Accordingly, we define the terms of average privacy and worst\hyph case privacy.

The formulation is then investigated theoretically.
Namely, we interpret a number of well\hyph known privacy criteria as particular cases of our more general metric.
The arguments behind these justifications are based on fundamental results related to the fields of information theory, probability theory and BDT.
More accurately, we interpret our privacy criterion as $k$\hyph anonymity and $l$\hyph diversity principles by connecting them to R\'enyi's entropy
and the concept of confidence set.
Under certain assumptions, a conditional version of the AEP allows us to interpret Shannon's entropy as an arbitrarily high confidence set.
%Another interpretation arises from Jaynes' rationale behind maximum entropy methods, which enables us to establish a connection to Shannon's entropy
%and KL divergence, both regarded as privacy measures in the case of user profiles.
Then, the total variation distance and Pinsker's inequality justify $t$\hyph closeness requirement and the criterion
proposed in~\cite{Rebollo10KDE} as particular instances of our measure of privacy.
In the course of this interpretation, we find that our formulation bears a strong resemblance with the rate\hyph distortion problem in
information theory.

Our survey of privacy metrics, our detailed analysis of their connection with information theory, and our mathematical unification as an attacker's estimation error, shed new light on the understanding of those metrics and their suitability when it comes to applying them to specific scenarios.
In regard to this aspect, two sections are devoted to the classification of several privacy metrics,
showing the relationships with our proposal and the correspondence with assumptions on the attacker's strategy.
While the former section approaches this from a theoretical perspective,
%the latter is written as a guide to help system designers choose the appropriate metrics,
the latter shows the applicability of our framework to those designers of SDC and ACSs who do not wish to delve into the mathematical details.
It is also our goal to illustrate the riveting interplay between the field of information privacy on the one hand, and on the other the fields of information theory and stochastic estimation, while bridging the gap between the respective communities.

A couple of simple albeit insightful examples are also presented.
Our first example quantifies the level of privacy provided by a privacy\hyph enhancing mechanism that perturbs location information in the scenario of LBS.
Under certain assumptions on the adversarial model, our measure of privacy becomes the mean squared error.
Then we turn our attention to the scenario of anonymous\hyph communication systems and
measure the degree of anonymity achieved by a modification of the collaborative protocol Crowds.
We contemplate different strategies for the attacker and, accordingly, interpret min\hyph entropy, Shannon's entropy and Hartley's entropy
as worst\hyph case, average\hyph case and best\hyph case privacy metrics.

%DAVID 111027: Rewritten
%Finally, we would like to remark that, although our theory is mature enough as it relies upon concepts from information theory and BDT,
%it is just a first step in the exploration of a unified privacy metric.
In closing, we hope that this unified perspective of privacy metrics, drawing upon the principles of
information theory and Bayesian estimation, is a helpful, illustrative step towards the systematic modeling
of privacy\hyph preserving information systems. 

\begin{table}[htb]
\centering
\includegraphics[scale=0.58]{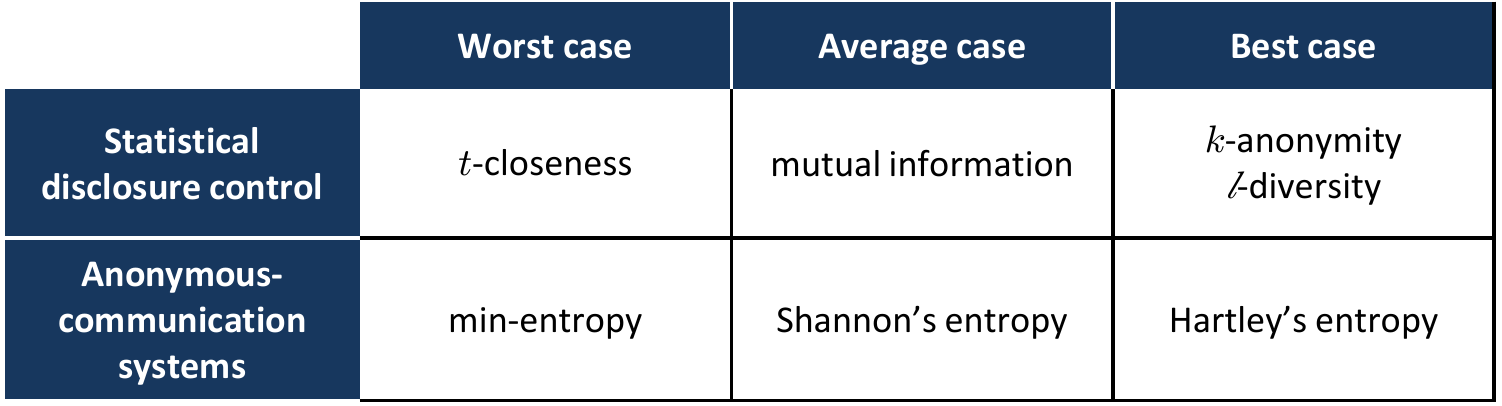}
\caption{Guide for designers of SDC and ACSs. This table classifies several privacy metrics depending, first, on whether they are regarded as worst\hyph case, average\hyph case and best\hyph case measures, and secondly on their application domain.}
\label{fig:Guide}
\end{table} 

%111018: Acknowledgment
%%%%%%%%%%%%%%%%%%%%%%%%%%%%%%%%%%%%%%%%%%%%%%%%%%%%%%%%%%%%%%%%%%%%%%%%%%%%%%%%%%%%%%%%%%%%%%%%%%%%%%%%%%%%%%%%%%%%%%%%%%%%%%%%%%
%%%%%%%%%%%%%%%%%%%%%%%%%%%%%%%%%%%%%%%%%%%%%%%%%%%%%%%%%%%%%%%%%%%%%%%%%%%%%%%%%%%%%%%%%%%%%%%%%%%%%%%%%%%%%%%%%%%%%%%%%%%%%%%%%%
%ACKNOWLEDGMENT
%%%%%%%%%%%%%%%%%%%%%%%%%%%%%%%%%%%%%%%%%%%%%%%%%%%%%%%%%%%%%%%%%%%%%%%%%%%%%%%%%%%%%%%%%%%%%%%%%%%%%%%%%%%%%%%%%%%%%%%%%%%%%%%%%%
%%%%%%%%%%%%%%%%%%%%%%%%%%%%%%%%%%%%%%%%%%%%%%%%%%%%%%%%%%%%%%%%%%%%%%%%%%%%%%%%%%%%%%%%%%%%%%%%%%%%%%%%%%%%%%%%%%%%%%%%%%%%%%%%%%
\section*{Acknowledgment}
\noindent
%We would like to thank
%the editor and the three anonymous referees for their thorough, extremely valuable comments,
%which motivated major improvements on this manuscript.
This work was partly supported by the Spanish Government through projects
Consolider Ingenio 2010 CSD2007-00004 ``ARES",
TEC2010-20572-C02-02 ``Consequence"
and by the Government of Catalonia under grant 2009 SGR 1362.
Additional sources of funding include IWT SBO SPION, GOA TENSE,
the IAP Programme P6/26 BCRYPT, and the FWO project ``Contextual privacy and the proliferation of location data".
D. Rebollo-Monedero is the recipient of a Juan de la Cierva postdoctoral fellowship,
JCI-2009-05259, from the Spanish Ministry of Science and Innovation.
C. Diaz is funded by an FWO postdoctoral grant.

%\bibliographystyle{IEEEtran}
%\bibliography{Bibliography/StringAbbreviated,Bibliography/Security,Bibliography/InfoTheory,Bibliography/LosslessCoding,Bibliography/LossyCoding,Bibliography/MathStatSigPro,Bibliography/Classification,Bibliography/Applications}

\end{document}